\documentclass[12pt, draftclsnofoot, onecolumn]{IEEEtran}
\IEEEoverridecommandlockouts
\usepackage{hyperref}
\usepackage{cite}
\usepackage{amsmath,amssymb,amsfonts,amsthm}
\usepackage{graphicx}
\usepackage{textcomp}
\usepackage[dvipsnames]{xcolor}
\usepackage{bm}
\usepackage{arydshln}
\usepackage{multirow}
\usepackage{mathdots}
\usepackage{mathtools}
\usepackage{setspace}
\usepackage{pgfplots}
\usepackage{tikz,pgfplots,tikzpeople}
\usetikzlibrary{positioning} 
\usepackage{nicematrix}
\usetikzlibrary{patterns}
\usepackage[labelformat=simple]{subcaption}
\usepackage{wasysym}
\usepackage{booktabs}
\usetikzlibrary{positioning,decorations.pathreplacing}

\newtheorem{theorem}{Theorem}

\newtheorem{lemma}{Lemma}
\newtheorem{remark}{Remark}

\let\emptyset\varnothing

\usepackage{algpseudocode,float}
\usepackage[ruled,linesnumbered]{algorithm2e}
\SetInd{0.5em}{0.5em}

\usepackage{lipsum}
\allowdisplaybreaks

\usepackage[bottom=1.08in, top=0.76in, left=0.64in, right=0.64in]{geometry}

\def\BibTeX{{\rm B\kern-.05em{\sc i\kern-.025em b}\kern-.08em
    T\kern-.1667em\lower.7ex\hbox{E}\kern-.125emX}}
    
\title{Robust Dynamic Coded Distributed Storage with Partially Storage Constrained Servers}

\author{
    \IEEEauthorblockN{Chen Zhao, Haobo Jia and Zhuqing Jia}\\
    \IEEEauthorblockA{School of Artificial Intelligence, Beijing University of Posts and Telecommunications \\
Email: \{chenzhao, jiahaobo, zhuqingj\}@bupt.edu.cn}
}

\begin{document}
\maketitle

\begin{abstract}
We consider the problem of Robust Dynamic Coded Distributed Storage (RDCDS) with partially storage constrained servers where the goal is to enable robust (resilient to server dropouts) and efficient (as measured by the communication costs) read and update operations, subject to the constraint that the storage at $S$ out of $N$ servers is limited by $1/K_c$ the size of the message. Building upon previously established converse arguments and achievability schemes by Jia et al., in this work we develop a set of new converse arguments and coding designs that enable us to completely characterize the fundamental limits of RDCDS with partially storage constrained servers, i.e., the minimum number of available servers for feasible update operation and the minimum communication costs for read and update operations across various server dropout scenarios.
\end{abstract}

\section{Introduction}
The problem of Robust Dynamic Coded Distributed Storage (RDCDS) addresses the challenge of distributing a message across $N$ servers to enable two key functionalities: robust and efficient data retrieval (read operation) and data modification (update operation). For the read operation, a user must be able to reconstruct the message by communicating with any subset of at least $R_r$ servers, thereby tolerating up to $N-R_r$ server dropouts, where $R_r$ is referred to as the read threshold. For the update operation, it allows a user to apply a local additive increment to the stored message. This is achieved by generating and uploading secure codewords representing the increment to a sufficient number of available servers. Specifically, the update requires communicating with at least $R_u^{(t)}$ available servers at time $t$, where $R_u^{(t)}$ is the update threshold. Besides, the uploaded codewords must be $X^{(t)}$-secured, i.e., any collusion of up to $X^{(t)}$ servers learns nothing about the increment itself. The superscript $t$ indicates the fact that the security level $X^{(t)}$ can be chosen arbitrarily (if feasible) by the user over time for each increment, and the update threshold $R_u^{(t)}$ is indeed a function of $X^{(t)}$. For the two operations, the key objective is to minimize the communication cost across various server dropout scenarios. Furthermore, both read and update operations must be memoryless, i.e., the user does not need knowledge of past server dropout states to perform current operations. The problem of RDCDS is related to various prominent research fields including codes for distributed storage \cite{Dimakis2006,dimakis2010network,Rashmi2011,Pawar2011,gopalan2012locality,Cadambe2013,Papailiopoulos2014,Song2014,Tamo2014,Guruswami2017,Goparaju2017,Ye2017repair,Ye2017secure,Rashmi2018,jia2024read,Li2024heterregen,Chou2024sds}, multi-version coding \cite{Wang2018}, communication-efficient secret sharing \cite{Wang2008secret,huang2016communication,bitar2017staircase,Martinez2018cess,Ding2022cess}, and can be seen as a special case of private read and write \cite{Sun2017pir,Sun2018tpir,Bita2018pir,jia2019xstpir,Jia_Jafar_MDSXSTPIR,Jia_Jafar_GXSTPIR,Jia2024Gxstplc,Jia_Jafar_XSTPFSL,Vithana2024commu,Vithana2023heter,Vithana2024contro,Vithana2024pfsl}. Furthermore, it may find applications to data aggregation within sensor networks \cite{Rajagopalan2006,wagner2004resilient,Roy2012sensor}, secure aggregation in federated learning \cite{bonawitz2017practical,kadhe2020fastsecagg,so2022lightsecagg,Wei2020fldp,Ulukus2022prcl}, etc. RDCDS under homogeneous storage constraint, where each of the $N$ servers stores no more than a $1/K_c$ fraction of the message, $K_c\in \mathbb{N}^*$, was explored in {\cite{jia2024read}}. It is established that the minimum update threshold is $R_u^{(t)} = N - R_r + K_c+ X^{(t)}$, the minimum normalized upload cost is $C_u^{(t)} = \frac{N-|\mathcal{D}^{(t)}|}{R_r - X^{(t)} - |\mathcal{D}^{(t)}|}$ and the minimum normalized download cost is $C_r^{(t)} = \frac{N-|\mathcal{D}^{(t)}|}{N - R_r + K_c - |\mathcal{D}^{(t)}|}$.

In this work, we investigate a more relaxed setting where only a subset of $S$ servers (out of $N$) is subject to the storage constraint of $1/K_c$, while the remaining $N-S$ servers have no explicit storage limit. This is termed as RDCDS with partially storage constrained servers. Our goal is to characterize the fundamental limits for this heterogeneous system, namely the minimum number of available servers for a feasible update ($R_u^{(t)}$), the minimum upload cost ($C_u^{(t)}$), and the minimum download cost ($C_r^{(t)}$) for the update and read operations, respectively. 
Leveraging converse bound arguments adapted from {\cite{jia2024read}}, we first establish lower bounds on the update threshold $R_u^{(t)}$, i.e., $R_u^{(t)} \ge N - R_r + X^{(t)} + S + 1$ is necessary when $S < K_c$ (i.e., the total storage capacity of the constrained servers is insufficient to hold the entire message, and this setting is non-trivial as with no unconstrained servers, the problem would have been infeasible), and $R_u^{(t)} \ge N - R_r + X^{(t)} +  K_c $ is necessary when $S \ge K_c$. Furthermore, converse bounds for the upload cost and the download cost are also established in the form of linear programming, where we develop a novel converse argument for the upload cost. Then, we show that the above bounds are tight by presenting matching achievability schemes, thus fully characterize the fundamental limits of RDCDS with partially storage constrained servers. Specifically, building upon the staircase code structure presented in {\cite{jia2024read}}, we introduce a novel re-encoding mechanism as the key component of our achievability scheme, which addresses the challenge posed by the partial storage constraints, i.e., staircase codewords initially designated for the servers that would exceed their $1/K_c$ limit are re-encoded and distributed to the servers with no storage constraints, with a carefully designed staircase code structure, while preserving the feasibility of both read and update operations. Although this re-encoding mechanism results in extra communication costs, it turns out that with the proposed overall coding structure, the lower bounds on communication costs and update threshold remain achievable. It is also notable that the extra storage cost required by this re-encoding mechanism results in a storage space of exactly $(1-\frac{S}{K_c})L$ when $S<K_c$, and $L/K_c$ when $S\ge K_c$ at those servers without storage constraints.

A preliminary version of this work, which omitted thorough proofs and technical details, was presented at the 2025 IEEE Information Theory Workshop (ITW). In contrast, this journal version provides a complete and rigorous treatment of our results. While the conference paper was limited to illustrating the key findings, this manuscript includes a strict and comprehensive proof and fully elaborates on the proposed achievability scheme for all possible scenarios. Furthermore, we have added a representative motivating example that clarifies the core insight and makes the result more accessible. By presenting the complete logical proofs and the detailed construction of our scheme, this article delivers a self-contained account of our research.

{\it Notations: }
Bold symbols are used to denote vectors and matrices, while calligraphic symbols denote sets. Following the convention, let the empty product be the multiplicative identity, and the empty sum be the additive identity. For two positive integers $M,N$ such that $M\le N,[M:N]$ denotes the set $\{M,M+1,\cdots,N\}$. We use the shorthand notation $[N]$ for $[1:N]$. $\mathbb{N}$ denotes the set of non-negative integers $\{0,1,2,\cdots\}$, and $\mathbb{N}^*$ denotes the set $\mathbb{N}\setminus\{0\}$. For a subset of integers $\mathcal{C}$, $\mathcal{C}(i),i\in[|\mathcal{C}|]$ denotes its $i^{th}$ element in ascending order. For a row (column) vector $\mathbf{v}$ of dimension $n$, $\mathbf{v}(\mathcal{I}), \mathcal{I}\subset[n]$ denotes the row (column) vector formed by the entries indexed by $\mathcal{I}$. We use the shorthand notation $\mathbf{v}(i)$ for $\mathbf{v}(\{i\})$, i.e., the $i^{th}$ entry of $\mathbf{v}$. For an $m\times n$ matrix $\mathbf{V}$ and two sets $\mathcal{A}\subset[m],\mathcal{B}\subset[n]$, $\mathbf{V}(\mathcal{A},\mathcal{B})$ denotes the submatrix of $\mathbf{V}$ formed by selecting rows indexed by $\mathcal{A}$ and columns indexed by $\mathcal{B}$. If $\mathcal{A}=[m]$ (or $\mathcal{B}=[n]$), it is abbreviated as the colon operator ($:$) in this context. We use the shorthand notation $\mathbf{V}(a,\mathcal{B}), \mathbf{V}(\mathcal{A},b), \mathbf{V}(a,b)$ for $\mathbf{V}(\{a\},\mathcal{B}),\mathbf{V}(\mathcal{A},\{b\}), \mathbf{V}(\{a\},\{b\})$ respectively. $\mathbf{0}_{m\times n}$ denotes the zero matrix of size $m\times n$.

\section{Problem Statement}  

\begin{figure}[!htbp]
    \centering
    \begin{subfigure}{1\columnwidth}
        \centering
\begin{tikzpicture}[xscale=0.6,yscale=0.8]

        \node [draw, rectangle,fill=orange!10, text=black, inner sep =0.1cm, rounded corners=0.5em] (S1) at (-9.5cm, 0cm) {\small\begin{tabular}{c}Server $1$ \\ $\mathbf{S}^{(t)}_{1}$\end{tabular}};

        \node [rectangle, inner sep =0.2cm, rounded corners=0.5em] (Ddots0) at (-7cm, 0cm) {$\cdots$};

        \node [draw, rectangle,fill=orange!10, text=black, inner sep =0.1cm, rounded corners=0.5em] (SS) at (-4.5cm, 0cm) {\small\begin{tabular}{c} Server $S$\\ $\mathbf{S}^{(t)}_{S}$\end{tabular}};

        \node [draw, rectangle,fill=teal!10, text=black, inner sep =0.1cm, rounded corners=0.5em] (SS1) at (-0.5cm, 0cm) {\small\begin{tabular}{c} Server $S+1$\\ $\mathbf{S}^{(t)}_{(S+1)}$\end{tabular}};

        \node [rectangle, inner sep =0.2cm, rounded corners=0.5em] (Ddots1) at (2.3cm, 0cm) {$\cdots$};

        \node [draw, rectangle,fill=teal!10, text=black, inner sep =0.1cm, rounded corners=0.5em] (S3) at (4.5cm, 0cm) {\small\begin{tabular}{c}Server $i$ \\ $\mathbf{S}^{(t)}_{i}$\end{tabular}};

        \node [rectangle, inner sep =0.2cm, rounded corners=0.5em] (Ddots2) at (7cm, 0cm) {$\cdots$};

        \node [draw, rectangle,fill=teal!10, text=black, inner sep =0.1cm, rounded corners=0.5em] (S4) at (9.5cm, 0cm) {\small\begin{tabular}{c}Server $N$ \\ $\mathbf{S}^{(t)}_{N}$\end{tabular}};

        \node[bob,minimum size=1cm] (User) at (0cm, -4cm) {User};
        
        \draw [black, thick, ->] (S1.south)to node [left=0.5cm] {\small $ \mathbf{A}_{1}^{(t)}$} (User);
        \draw [black, thick, ->] (SS.south)to node [left=0.2cm] {\small $ \mathbf{A}_{S}^{(t)}$} (User);
        \draw [black, thick, ->] (SS1.south)to node [left=0cm] {\small $ \mathbf{A}_{(S+1)}^{(t)}$} (User);
        
        \draw [red, dashed, ->] (S3.south)to node [left=0.1cm] {\large $\varnothing$} (User);
        \draw [black, thick, ->] (S4.south)to node [left=0.5cm] {\small $ \mathbf{A}_{N}^{(t)}$} (User);

        \node[below=0.2cm of User] (bk){};
    
    \end{tikzpicture}
    \caption{The read operation is executed at time slot $t$, where Server $i$ is unavailable, i.e., $i\in\mathcal{D}^{(t)}$.}
    \vspace{0.5cm}
    \end{subfigure}
    
    \begin{subfigure}{1\columnwidth}
        \centering
        \begin{tikzpicture}[xscale=0.6,yscale=0.8]

        \node[bob,minimum size=1cm] (User) at (0cm, -4cm) {User};
        \node[below=0.2cm of User] (bk){};

        \node [draw, rectangle,fill=orange!10, text=black, inner sep =0.1cm, rounded corners=0.5em] (S1) at (-9.5cm, 0cm) {\small\begin{tabular}{c}Server $1$ \\ $\mathbf{S}^{(t+1)}_{1}$\\$\uparrow$\\$\mathbf{S}^{(t)}_{1}$\end{tabular}};

        \node [rectangle, inner sep =0.1cm, rounded corners=0.5em] (Ddots0) at (-7cm, 0cm) {$\cdots$};
        
        \node [draw, rectangle,fill=orange!10, text=black, inner sep =0.1cm, rounded corners=0.5em] (SS) at (-4.5cm, 0cm) {\small\begin{tabular}{c} Server $S$\\ $\mathbf{S}^{(t+1)}_{S}$\\$\uparrow$\\ $\mathbf{S}^{(t)}_{S}$\end{tabular}};

        \node [draw, rectangle,fill=teal!10, text=black, inner sep =0.1cm, rounded corners=0.5em] (SS1) at (-0.5cm, 0cm) {\small\begin{tabular}{c} Server $S+1$\\ $\mathbf{S}^{(t+1)}_{(S+1)}$\\$\uparrow$\\ $\mathbf{S}^{(t)}_{(S+1)}$\end{tabular}};

        \node [rectangle, inner sep =0.2cm, rounded corners=0.5em] (Ddots1) at (2.3cm, 0cm) {$\cdots$};

        \node [draw, rectangle,fill=teal!10, text=black, inner sep =0.1cm, rounded corners=0.5em] (S3) at (4.5cm, 0cm) {\small\begin{tabular}{c}Server $i$\\ $\mathbf{S}^{(t+1)}_{i}$\\$\parallel$ \\ $\mathbf{S}^{(t)}_{i}$\end{tabular}};

        \node [rectangle, inner sep =0.1cm, rounded corners=0.5em] (Ddots2) at (7cm, 0cm) {$\cdots$};

        \node [draw, rectangle,fill=teal!10, text=black, inner sep =0.1cm, rounded corners=0.5em] (S4) at (9.5cm, 0cm) {\small\begin{tabular}{c}Server $N$\\ $\mathbf{S}^{(t+1)}_{N}$\\$\uparrow$ \\ $\mathbf{S}^{(t)}_{N}$\end{tabular}};
                
        \node[minimum size=0.3cm, inner sep=0.1cm] (W) at (0cm, -6cm) {$\boldsymbol{\Delta}^{(t)}$};
        \draw [black, thick, ->] (W)--(bk);

        \draw [black, thick,  ->] (User)to node [left=0.5cm] {\small $ \mathbf{Q}_{1}^{(t)}$}(S1.south);
        \draw [black, thick,  ->] (User)to node [left=0.2cm] {\small $ \mathbf{Q}_{S}^{(t)}$}(SS.south); 
        
        \draw [black, thick,  ->] (User)to node [right=0cm] {\small $ \mathbf{Q}_{(S+1)}^{(t)}$}(SS1.south); 
        \draw [red, dashed,  ->] (User)to node [right=0.1cm] {\large $\varnothing$}(S3.south);
        \draw [black, thick,  ->] (User)to node [right=0.5cm] {\small $ \mathbf{Q}_{N}^{(t)}$}(S4.south);
   
    \end{tikzpicture}
        \caption{The update operation is executed at time slot $t$, where Server $i$ is unavailable, i.e, $i\in\mathcal{D}^{(t)}$.
        }

    \end{subfigure}
    
    \caption{The problem of robust dynamic coded distributed storage with partially storage constrained servers. The servers marked with light orange have storage limits; the others have no explicit storage limit.}
    \label{fig:PRDCDS}
\end{figure}
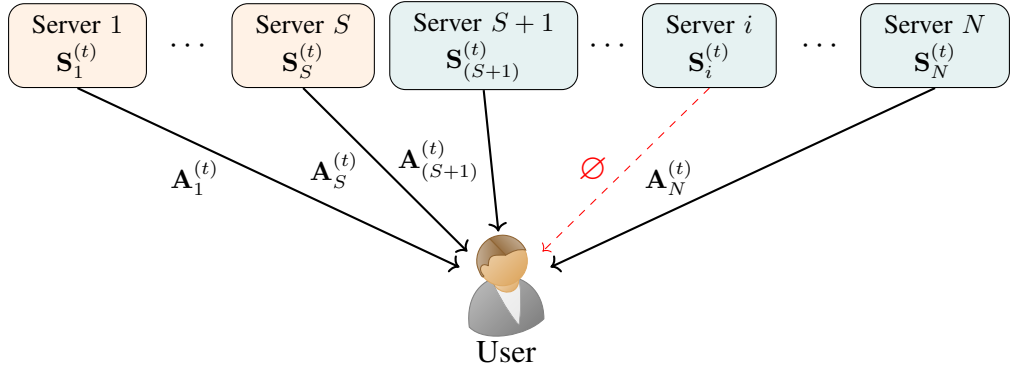
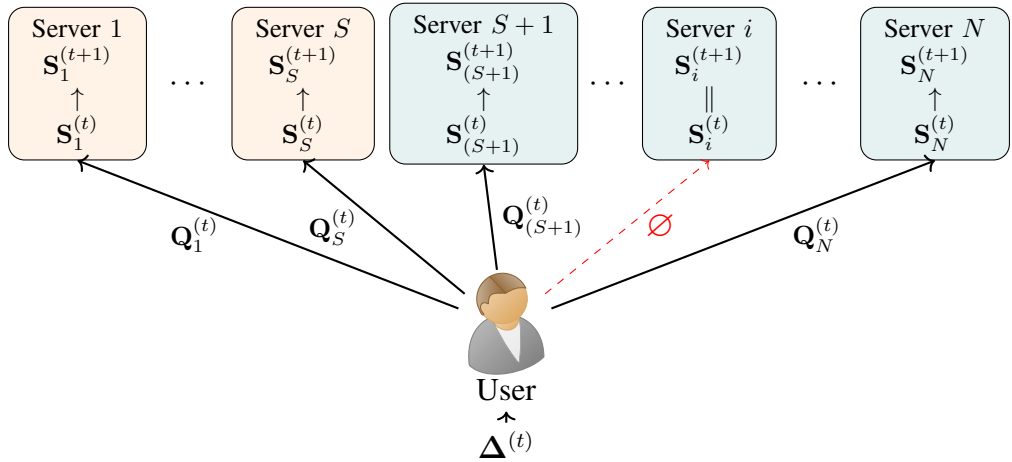

Consider a distributed storage system of $N$ servers. As shown in Figure \ref{fig:PRDCDS}, the problem is associated with the coded distributed storage of a message over time slots $t$, i.e., for all $t\in \mathbb{N}$, the overall storage must form a $(K_c,R_r,N,S)$ (partially constrained) coded distributed storage of the message $\mathbf{W}^{(t)}$ that consists of $L$ (i.i.d.) symbols from the finite field $\mathbb{F}_q$.
Without loss of generality, let us assume that the set of storage constrained servers is $[S]$, i.e., the first $S$ servers. Formally, denoting the storage at Server $n$ as $\mathbf{S}_n^{(t)}$, $\forall t\in\mathbb{N}$, we must have
\begin{itemize}
    \item {\bf $R_r$-recoverability: }The message must be a deterministic function of the storage at any $R_r$ servers, i.e., $\forall\mathcal{R}\subset [N]$ such that $|\mathcal{R}|=R_r$,
    \begin{align}
        H(\mathbf{W}^{(t)}\mid (\mathbf{S}_n^{(t)})_{n\in\mathcal{R}})=0.\label{def:Rr-rec}
    \end{align}
    \item {\bf $(K_c,S)$-storage cost: }The storage at storage constrained server is at most $\frac{1}{K_c}$ the size of the message, i.e., $\forall n \in [S]$,
    \begin{align}
        H(\mathbf{S}_n^{(t)})\leq \frac{L}{K_c},\label{eq:minstor}
    \end{align}
    in $q$-ary units.
\end{itemize}

 Recall that when $S<K_c$, the total storage capacity of the constrained servers is insufficient to hold the entire message, therefore $R_r$-recoverability requires $S<R_r$ for this setting. Besides, when $t=0$, we assume that the initial storage $(\mathbf{S}_n^{(0)})_{n\in[N]}$ and the corresponding message $\mathbf{W}^{(0)}$ is initialized {\it a priori} by a global coordinator.

Suppose at any time slot $t$, $t\in\mathbb{N}^*$, the user wishes to execute either the read operation or the update operation. A subset of servers may be temporarily inaccessible, which are referred to as dropout servers, denoted as $\mathcal{D}^{(t)}$. The dropout servers may be randomly drawn, but once realized, are considered to be constant during the time slot and globally known to both the (available) servers and the user. If the read operation is executed at time slot $t$, the user downloads symbols from the available servers $[N]\setminus\mathcal{D}^{(t)}$ to recover the message $\mathbf{W}^{(t)}$. Denoting downloaded symbols from Server $n$ as $\mathbf{A}_n^{(t)}$, we have
\begin{itemize}
    \item {\bf Determinacy: }The downloaded symbols from each of the servers must be a deterministic function of its storage, i.e., for all $n\in[N]\setminus\mathcal{D}^{(t)}$,
    \begin{align}
        H(\mathbf{A}_n^{(t)}\mid \mathbf{S}_n^{(t)})=0.\label{def:detmin}
    \end{align}
    \item {\bf Correctness: }The message must be recoverable from the downloads, i.e.,
    \begin{align}
        H(\mathbf{W}^{(t)}\mid (\mathbf{A}_{n}^{(t)})_{n\in [N]\setminus\mathcal{D}^{(t)}})=0.\label{def:dlcrec}
    \end{align}
    \item {\bf Storage transition: }The storage at each server is untouched by the read operation, i.e., for all $n\in \mathcal{D}^{(t)}$,
    \begin{align}\label{def:stortranrd}
        \mathbf{S}_n^{(t+1)}=\mathbf{S}_n^{(t)}
    \end{align}
    and
    \begin{align}\label{def:wtranrd}
        \mathbf{W}^{(t+1)}=\mathbf{W}^{(t)}.
    \end{align}

\end{itemize}

The communication efficiency of the read operation is characterized by the normalized download cost $C_r^{(t)}$, defined as 
\begin{align}
    C_r^{(t)}=\frac{\sum_{n\in [N]\setminus\mathcal{D}^{(t)}} H(\mathbf{A}_n^{(t)})}{L}.
\end{align}

At time slot $t$, if the update operation is executed, the message is updated by the local additive increment generated by the user $\boldsymbol{\Delta}^{(t)}$ consisting $L$ (i.i.d.) symbols from the finite field $\mathbb{F}_q$. To this end, each of the available servers updates its storage according to the $X^{(t)}$-securely coded increment uploaded by the user, denoted as $\mathbf{Q}_n^{(t)}, n\in[N]\setminus\mathcal{D}^{(t)}$, such that

\begin{itemize}
    \item {\bf Correctness: }The message must be additively updated by the increment, i.e.,
    \begin{align}
        \mathbf{W}^{(t+1)}=\mathbf{W}^{(t)}+\boldsymbol{\Delta}^{(t)}.\label{def:up-correct}
    \end{align}
    \item {\bf $X^{(t)}$-security: }The increment must be independent of any $X^{(t)}$ coded increments, $0\leq X^{(t)}< R_r$, i.e., for all $\mathcal{X}\subset [N]\setminus\mathcal{D}^{(t)}$ such that $|\mathcal{X}|=X^{(t)}$,
    \begin{align}
        I(\boldsymbol{\Delta}^{(t)}; (\mathbf{Q}_n^{(t)})_{n\in\mathcal{X}})=0.\label{def:Xsec}
    \end{align}
    \item {\bf Storage transition: }The updated storage at each of the available servers must be a deterministic function of the current storage and the coded increment, i.e., for all $n\in [N]\setminus\mathcal{D}^{(t)}$,
    \begin{align}
        H(\mathbf{S}_n^{(t+1)}\mid \mathbf{S}_n^{(t)}, \mathbf{Q}_n^{(t)})=0.\label{def:stotrans}
    \end{align}
    As for the dropout servers, their storage must left untouched, i.e., for all $n\in\mathcal{D}^{(t)}$,
    \begin{align}\label{eq:uddropstor}
        \mathbf{S}_n^{(t+1)}=\mathbf{S}_n^{(t)}.
    \end{align}
    \item {\bf Independence: }The increment is independent of the current message, and the user has no prior information on the server storage, i.e.,
    \begin{align}
        I(\boldsymbol{\Delta}^{(t)},(\mathbf{Q}_n^{(t)})_{n\in[N]\setminus\mathcal{D}^{(t)}};(\mathbf{S}_n^{(t)})_{n\in[N]})=0.\label{def:ind}
    \end{align}

\end{itemize}

The update threshold $R_u^{(t)}$ is defined as the minimum number of available servers required by the scheme so that the system remains operational after the update operation as well as guaranteeing $X^{(t)}$-security during the process, i.e., the storage at time $t+1$ remains $(K_c, R_r,N,S)$-coded distributed storage of $\mathbf{W}^{(t+1)}$  regardless of the realization of dropout servers $\mathcal{D}^{(t)}$. Note that the updated message must be recoverable by the user at time slot $t+1$ by communicating with any $R_r$ servers, therefore we must have $0\leq X^{(t)}< R_r$, otherwise, any $R_r$ servers have no information on the increment, hence the updated message. The communication efficiency of the update operation is characterized by the normalized upload cost $C_u^{(t)}$, defined as 
\begin{align}
    C_u^{(t)}=\frac{\sum_{n\in [N]\setminus\mathcal{D}^{(t)}} H(\mathbf{Q}_n^{(t)})}{L}.
\end{align} 

As {\it robust} dynamic coded distributed storage requires memoryless read and update operations, the user is indeed oblivious to the time index $t$, i.e., there is no need to know how many rounds of read or update operations have been performed before. Therefore, we are particularly interested in the setting where the system is in its steady state, i.e., $t>t_0$ for some sufficiently large $t_0$, where the system must have experienced all possible dropout states, increment security levels, etc.

\section{Main Results}
The main result of this work is the complete characterization of the fundamental limits of RDCDS with partially storage constrained servers, formally presented in the following theorem. To make the result more concise, let us define 
\[
\Omega = 
\begin{cases}
R_r - S - 1, & \text{if } S < K_c, \\
R_r - K_c,   & \text{if } S \ge K_c.
\end{cases}
\]
\begin{theorem}\label{thm:main}
    At time slot $t>t_0$, we have 
    \begin{align}
        &R_u^{(t)}\geq N-\Omega+X^{(t)},\\ &C_r^{(t)}\geq\min_{(D_1, \cdots, D_N)\in\mathcal{D}}\sum_{n\in[N]\setminus\mathcal{D}^{(t)}}D_n,\\
        &C_u^{(t)}\geq\min_{(U_1, \cdots, U_N)\in\mathcal{U}}\sum_{n\in[N]\setminus\mathcal{D}^{(t)}}U_n, 
    \end{align}where $\mathcal{D}$ is the set of $N$-dimensional vectors $(D_1, \cdots, D_N)\in\mathbb{R}_+^N$ such that
    \begin{align}
        \textstyle\sum_{n\in[N]\setminus\mathcal{X}\setminus\mathcal{D}^{(t)}}D_n\geq 1,& ~\forall\mathcal{X}\subset[N]\setminus\mathcal{D}^{(t)},|\mathcal{X}|=\Omega,\label{eq:defD-1}\\
        D_n\leq 1/K_c,&~\forall n\in[S]\setminus \mathcal{D}^{(t)}\label{eq:defD},
    \end{align}
    and $\mathcal{U}$ is the set of $N$-dimensional vectors $(U_1, \cdots, U_N)\in\mathbb{R}_+^N$ such that
    \begin{align}
        \textstyle\sum_{n\in\mathcal{R}\setminus\mathcal{X}}U_n\geq 1,&~\forall\mathcal{R}\subset[N]\setminus\mathcal{D}^{(t)},|\mathcal{R}|=R_r-|\mathcal{D}^{(t)}|,\notag\\
        \quad \quad& \quad\quad\quad\quad\quad\quad\mathcal{X}\subset\mathcal{R},|\mathcal{X}|=X^{(t)}\label{eq:defU-1},\\
        U_n\leq 1/K_c,&~\forall n\in[S]\setminus \mathcal{D}^{(t)}\label{eq:defU}.
    \end{align}
    And our scheme achieves the above bounds simultaneously at any time slot $t\in\mathbb{N}$. 
\end{theorem}
\begin{remark}\label{remark:homogeneous}
As mentioned in the introduction section, when $S\geq K_c$, the update threshold $R_u^{(t)}\geq N-R_r+X^{(t)}+ K_c$, which coincides with the result of the homogeneous setting in {\cite{jia2024read}}. Besides, averaging over all possible $\mathcal{X}$, we have $C_r^{(t)} \geq \frac{N-|\mathcal{D}^{(t)}|}{N - R_r + K_c - |\mathcal{D}^{(t)}|}$ according to the first bound in \eqref{eq:defD}, and averaging over all possible $\mathcal{X},\mathcal{R}$ in the first bound of \eqref{eq:defU} we have $C_u^{(t)} \geq \frac{N-|\mathcal{D}^{(t)}|}{R_r - X^{(t)} - |\mathcal{D}^{(t)}|}$. Since the above bounds are achievable by the homogeneous scheme in {\cite{jia2024read}}, this setting simply reduces to the homogeneous case. Therefore, in the achievability proof of Theorem \ref{thm:main}, we mainly focus on the non-trivial setting where $S< K_c$.
\end{remark}

\section{Converse Proof of Theorem \ref{thm:main}}
The following lemmas due to {\cite{jia2024read}} are necessary for our proof, which are reproduced for completeness.
\begin{lemma}\label{lemma:rdcds1}
    Assume that the update operation is executed at time slot $t$. Then 
    for all $\mathcal{R}, \mathcal{X}$ such that $\mathcal{D}^{(t)}\subset \mathcal{R}\subset[N]$, $\mathcal{X}\subset\mathcal{R}\setminus\mathcal{D}^{(t)}$, $|\mathcal{R}|=R_r, |\mathcal{X}|=X^{(t)}$, we have $\sum_{n\in\mathcal{R}\setminus\mathcal{X}\setminus\mathcal{D}^{(t)}}H(\mathbf{S}_n^{(t+1)})\ge L$.
\end{lemma}

\begin{lemma}\label{lemma:indeinh}
    Assume that the update operation is executed at time slot $t$. Then for all $\mathcal{X}\subset[N]$ such that $|\mathcal{X}|=X^{(t)}, \mathcal{X}\cap\mathcal{D}^{(t)}=\emptyset$ and all $\tau\in\mathbb{N}^*$, we have $I\left((\mathbf{S}_n^{(t+\tau)})_{n\in\mathcal{X}\cup\mathcal{D}^{(t)}}; \mathbf{W}^{(t+\tau)}\right)=0$.
\end{lemma}

\begin{lemma}\label{lemma:rdcds3}
    Assume that the update operation is executed at time slot $t$. Then for all $\mathcal{X},\mathcal{R}$ such that $\mathcal{X}\subset\mathcal{R}\subset[N]\setminus\mathcal{D}^{(t)}$ and $|\mathcal{X}|=X^{(t)}, |\mathcal{R}|=R_r-|\mathcal{D}^{(t)}|$, we have
    $\sum_{n\in\mathcal{R}\setminus\mathcal{X}} H(\mathbf{Q}_n^{(t)})\ge L$.
\end{lemma}

Now we are ready to formally prove the converse bounds in Theorem \ref{thm:main}.
\begin{proof}{\bf (Lower bound on the update threshold $R_u^{(t)}$)}
    Let us assume that the update operation is executed at time slot $t$. According to Lemma \ref{lemma:rdcds1}, for all $\mathcal{R}, \mathcal{X}$ such that $\mathcal{R}\supset[S],\mathcal{D}^{(t)}\subset \mathcal{R}\subset[N],|\mathcal{R}|=R_r$, $\mathcal{X}\subset\mathcal{R}\setminus\mathcal{D}^{(t)}$, $ |\mathcal{X}|=X^{(t)}$, we have 
    \begin{align}
    \textstyle\sum_{n\in\mathcal{R}\setminus\mathcal{X}\setminus\mathcal{D}^{(t)}}H(\mathbf{S}_n^{(t+1)})\ge L\label{eq:base}.
    \end{align}
    
    By the definition of the update threshold, our scheme should remain operational regardless of the realization of $\mathcal{D}^{(t)}$. That means the scheme should tolerate a certain size of the dropout servers occurring in all possible situations. Let us establish a proof by contradiction to validate the lower bound of $R_u^{(t)}$.
    
    For the setting where $S<K_c$, {assume dropout occurs among Server $S+1$, Server $S+2$, $\cdots$, Server $N$ such that $|\mathcal{D}^{(t)}|=R_r-X^{(t)}-S$,} which happen to violate the update threshold $R_u^{(t)}\geq N-R_r+X^{(t)}+S+1$. Now note that we can fix sets $\mathcal{R},\mathcal{X}$  such that $\mathcal{D}^{(t)}\subset \mathcal{R}\subset[N]$, $\mathcal{X}\subset\mathcal{R}\setminus\mathcal{D}^{(t)}$, $|\mathcal{R}|=R_r, |\mathcal{X}|=X^{(t)}$ and $\mathcal{R}\setminus\mathcal{X}\setminus\mathcal{D}^{(t)}=[S]$, as shown in Fig.\ref{fig:Venn-1}. Now it is obvious that $\sum_{n\in\mathcal{R}\setminus\mathcal{X}\setminus\mathcal{D}^{(t)}}H(\mathbf{S}_n^{(t+1)})=\sum_{n\in[S]}H(\mathbf{S}_n^{(t+1)})=\frac{S}{K_c}L<L$. It contradicts \eqref{eq:base} and thus we have $R_u^{(t)}\geq N-R_r+X^{(t)}+S+1$.
        
    Similarly, for the setting where $S\ge K_c$, assume the number of dropout servers is $|\mathcal{D}^{(t)}| = R_r - X^{(t)} - K_c + 1$, and at least $K_c - 1$ servers in the set $[S]$ are available. It violates the update threshold $R_u^{(t)}\geq N-R_r+X^{(t)}+K_c$ as well. Now, if we fix sets $\mathcal{R},\mathcal{X},\mathcal{S}'$ such that $\mathcal{D}^{(t)}\subset \mathcal{R}\subset[N]$, $\mathcal{X}\subset\mathcal{R}\setminus\mathcal{D}^{(t)}$, $\mathcal{S}'=\mathcal{R}\setminus\mathcal{X}\setminus\mathcal{D}^{(t)}\subset[S]$,  and $|\mathcal{R}|=R_r, |\mathcal{X}|=X^{(t)}$, $|\mathcal{S}'|=K_c-1$, as shown in Fig.\ref{fig:Venn-2}. Then we have $\sum_{n\in\mathcal{R}\setminus\mathcal{X}\setminus\mathcal{D}^{(t)}}H(\mathbf{S}_n^{(t+1)})=\sum_{n\in\mathcal{S}'}H(\mathbf{S}_n^{(t+1)})=\frac{K_c-1}{K_c}L<L$, which contradicts \eqref{eq:base}. Therefore, we must have $R_u^{(t)}\ge N-R_r+X^{(t)}+ K_c$.
    
    \begin{figure}

        \begin{subfigure}{1\columnwidth}
               \centering
        \begin{tikzpicture}[font=\small, scale=1.2, every node/.style={align=center}]
            
            
            
    
            \draw[thick] (0,-2) rectangle (10,-3);
            
            \draw[fill=blue!10, thick] (0,-2) rectangle (4,-3);
            \node at (2,-2.5) {$[S]$};
            
            \draw[fill=green!20, thick] (4,-2) rectangle (5,-3);
            \node at (4.5,-2.5) {$\mathcal{X}$};
            \draw[decorate, decoration={brace, amplitude=10pt}] 
              (4,-2) -- (5,-2) node[midway, above=10pt]{$|\mathcal{X}|=X^{(t)}$};
    
            \draw[fill=orange!30, thick] (5,-2) rectangle (7,-3);
            \node at (6,-2.5) {$\mathcal{D}^{(t)}$};
            \draw[decorate, decoration={brace, amplitude=10pt}] 
              (5,-2) -- (7,-2) node[above=10pt]{$|\mathcal{D}^{(t)}|=R_r-X^{(t)}-S$};
              
            \draw[decorate, decoration={brace, mirror, amplitude=10pt}] 
              (0,-3) -- (7,-3) node[midway, below=10pt]{$\mathcal{R},|\mathcal{R}|=R_r$};

            \node at (8.5, -2.5) {$[N]\setminus\mathcal{R}$};
        
        \end{tikzpicture}
        \caption{The realization of the sets $\mathcal{R},\mathcal{X}$ for the setting $S<K_c$.}
        \label{fig:Venn-1}
        \end{subfigure}

        \vspace{0.5 cm}
        
        \begin{subfigure}{1\columnwidth}
            \centering
        \begin{tikzpicture}[font=\small, scale=1.2, every node/.style={align=center}]
            
            
            
    
            \draw[thick] (0,-2) rectangle (10,-3);
            
            \draw[fill=blue!20, thick] (0,-2) rectangle (3,-3);
            \node at (1.5,-2.5) {$\mathcal{S}',\mathcal{S}'\subset[S]$};
            
            \draw[fill=green!20, thick] (3,-2) rectangle (4,-3);
            \node at (3.5,-2.5) {$\mathcal{X}$};
            \draw[decorate, decoration={brace, amplitude=10pt}] 
              (3,-2) -- (4,-2) node[pos=0.5, above=10pt]{$|\mathcal{X}|=X^{(t)}$};
    
            \draw[fill=orange!30, thick] (4,-2) rectangle (7,-3);
            \node at (5.5,-2.5) {$\mathcal{D}^{(t)}$};
            \draw[decorate, decoration={brace, amplitude=10pt}] 
              (4,-2) -- (7,-2) node[pos=0.8, above=10pt]{$|\mathcal{D}^{(t)}|=R_r-X^{(t)}-K_c+1$};
              
            \draw[decorate, decoration={brace, mirror, amplitude=10pt}] 
              (0,-3) -- (7,-3) node[midway, below=10pt]{$\mathcal{R},|\mathcal{R}|=R_r$};

            \node at (8.5, -2.5) {$[N]\setminus\mathcal{R}$};
        
        \end{tikzpicture}
        \caption{The realization of the sets $\mathcal{R},\mathcal{X},\mathcal{S}'$ for the setting $S\geq K_c$.}
        \label{fig:Venn-2}
        \end{subfigure}
        \caption{The illustration of the realization of the corresponding sets in the proof of the lower bound on the update threshold $R_u^{(t)}$.}
    \end{figure}
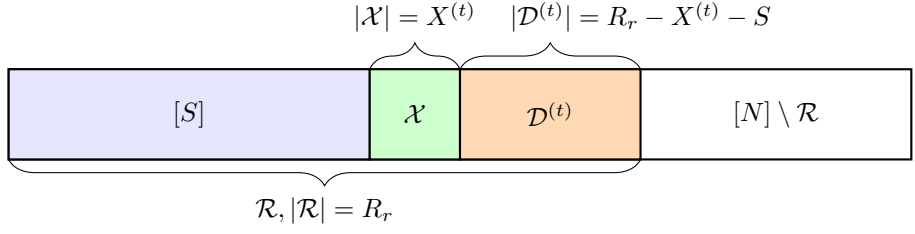
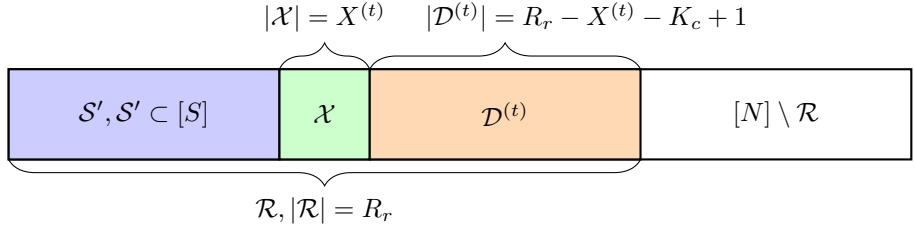

{\bf (Lower bound on the download cost $C_r^{(t)}$)}
    For all $n\in [S]\setminus\mathcal{D}^{(t)}$, due to $(K_c,S)$-storage constraint and determinacy, we have $\frac{H(\mathbf{A}_n^{(t)})}{L}\leq\frac{H(\mathbf{S}_n^{(t)})}{L}\leq \frac{1}{K_c}$. On the other hand, according to the lower bound on $R_u^{(t)}$, we have {$|\mathcal{D}^{(t)}|\leq \Omega-X^{(t)}$} for all $t\in\mathbb{N}^*$, let us suppose the user executes the update operation at various time slots $t_1, t_2, \cdots$ and the sets $\mathcal{D}^{(t_1)}\cup\mathcal{X}^{(t_1)}, \mathcal{D}^{(t_2)}\cup\mathcal{X}^{(t_2)},\cdots$ take all possible subsets of $[N]$ such that $|\mathcal{D}^{(t_i)}\cup\mathcal{X}^{(t_i)}|=\Omega$, where $\mathcal{X}^{(t_i)}$ is an arbitrary subset of $[N]$ such that $\mathcal{X}^{(t_i)}\cap\mathcal{D}^{(t_i)}=\emptyset,|\mathcal{X}^{(t_i)}|=X^{(t_i)}$. 
    Then according to Lemma \ref{lemma:indeinh}, there must exist some sufficient large $t_0\in\mathbb{N}^*$ such that for all $t>t_0$ and $\mathcal{X}\subset[N]\setminus\mathcal{D}^{(t)},|\mathcal{X}|=\Omega$, we have
    \begin{align}\label{eq:scxsec}
        I\left((\mathbf{S}_n^{(t)})_{n\in\mathcal{X}}; \mathbf{W}^{(t)}\right)=0. 
    \end{align}
    Thus, we have
    \begin{align}
           L&=H(\mathbf{W}^{(t)})\\
            &=I(\mathbf{W}^{(t)};(\mathbf{A}_n^{(t)})_{n\in[N]\setminus\mathcal{D}^{(t)}})+H(\mathbf{W}^{(t)}\mid (\mathbf{A}_n^{(t)})_{n\in[N]\setminus\mathcal{D}^{(t)}})\label{eq:lbdc-1}\\
            &=I(\mathbf{W}^{(t)};(\mathbf{A}_n^{(t)})_{n\in[N]\setminus\mathcal{D}^{(t)}})\label{eq:lbdc-2}\\
            &=I(\mathbf{W}^{(t)};(\mathbf{A}_n^{(t)})_{n\in\mathcal{X}})+I(\mathbf{W}^{(t)};(\mathbf{A}_n^{(t)})_{n\in[N]\setminus\mathcal{X}\setminus\mathcal{D}^{(t)}}\mid (\mathbf{A}_n^{(t)})_{n\in\mathcal{X}})\label{eq:lbdc-3}\\
            &\leq I(\mathbf{W}^{(t)};(\mathbf{S}_n^{(t)})_{n\in\mathcal{X}})+I(\mathbf{W}^{(t)};(\mathbf{A}_n^{(t)})_{n\in[N]\setminus\mathcal{X}\setminus\mathcal{D}^{(t)}}\mid (\mathbf{A}_n^{(t)})_{n\in\mathcal{X}})\label{eq:lbdc-4}\\
            &=I(\mathbf{W}^{(t)};(\mathbf{A}_n^{(t)})_{n\in[N]\setminus\mathcal{X}\setminus\mathcal{D}^{(t)}}\mid (\mathbf{A}_n^{(t)})_{n\in\mathcal{X}})\label{eq:lbdc-5}\\
            &\leq H\left((\mathbf{A}_n^{(t)})_{n\in[N]\setminus\mathcal{X}\setminus\mathcal{D}^{(t)}}\mid (\mathbf{A}_n^{(t)})_{n\in\mathcal{X}}\right)\label{eq:lbdc-6}\\
            &\leq H\left((\mathbf{A}_n^{(t)})_{n\in[N]\setminus\mathcal{X}\setminus\mathcal{D}^{(t)}}\right)\label{eq:lbdc-7}\\
            &\leq \sum_{n\in[N]\setminus\mathcal{X}\setminus\mathcal{D}^{(t)}}H(\mathbf{A}_n^{(t)})\label{eq:lbdc-8}
    \end{align}
        in $q$-ary units, where \eqref{eq:lbdc-2} holds due to the correctness constraint in read operation \eqref{def:dlcrec}. \eqref{eq:lbdc-4} holds because the determinacy constraint \eqref{def:detmin}, i.e., $\mathbf{A}_n^{(t)}$ is fully determined by $\mathbf{S}_n^{(t)}$ for all $n\in \mathcal{X}\subset[N]\setminus{\mathcal{D}^{(t)}}$. And \eqref{eq:lbdc-5} holds due to \eqref{eq:scxsec}.
     Combining the above bounds, we obtain the desired bound $C_r^{(t)}\geq\min_{(D_1, \cdots, D_N)\in\mathcal{D}}\sum_{n\in[N]}D_n$, where $\mathcal{D}$ is defined in \eqref{eq:defD}.

{\bf (Lower bound on the upload cost $C_u^{(t)}$)}
    Consider a relaxed setting where at any time slot $t$, the storage $(\mathbf{S}^{(t)}_n)_{n\in[N]}$ is made known to the user. For this setting, we note that if the update operation is executed at time slot $t$, there is no loss of generality by assuming that the normalized upload cost {for each of the available, storage-constrained servers} is upper bounded by $1/K_c$. This is because $\mathbf{S}^{(t+1)}_n$ is a deterministic function of $\mathbf{S}^{(t)}_n$ and coded increment. Since $\mathbf{S}^{(t)}_n$ is made globally known, the user can always generate and upload $\mathbf{S}^{(t+1)}_n$ trivially, and the bound follows. 
    
    On the other hand, since $\mathbf{S}^{(t)}_n$ is independent of the increment, the bounds in Lemma \ref{lemma:rdcds3} still hold, i.e., in the worst case, one must be able to recover the increment from its coded version for any set of servers $\mathcal{R}\setminus\mathcal{X}$ such that $\mathcal{X}\subset\mathcal{R}\subset[N]\setminus\mathcal{D}^{(t)}$ and $|\mathcal{X}|=X^{(t)}, |\mathcal{R}|=R_r-|\mathcal{D}^{(t)}|$. 
    
    Combining the above bounds, for this relaxed setting, the minimum normalized upload cost $\overline{C}_u^{(t)}$ is then lower bounded by $\overline{C}_u^{(t)}\geq \min_{(U_1, \cdots, U_N)\in\mathcal{U}}\sum_{n\in[N]}U_n$, where $\mathcal{U}$ is defined in \eqref{eq:defU}. {Now note that the relaxing problem cannot hurt the upload cost, we must have ${C}_u^{(t)}\geq\overline{C}_u^{(t)}\geq \min_{(U_1, \cdots, U_N)\in\mathcal{U}}\sum_{n\in[N]}U_n$.}
\end{proof}

\section{Achievability Proof of Theorem \ref{thm:main}}
To make our achievability scheme more accessible, let us start with a motivating example. Besides, in order to distinguish between the dropouts of storage-unconstrained servers and those of storage-constrained servers, let us define $\mathcal{D}_1^{(t)}=[S+1:N]\cap\mathcal{D}^{(t)}$ and $\mathcal{D}_2^{(t)}=[S]\cap\mathcal{D}^{(t)}$.

\subsection{Motivating Example}
We consider the setting where $N=7,R_r=5,K_c=6,S=2$, i.e., the overall storage of the $N$ servers forms $(K_c,R_r,N,S)=(6,5,7,2)$ partially constrained coded distributed storage at any time slot $t, t\in \mathbb{N}$. 

\subsubsection{Construction of the Storage}
 Let $x_1,x_2,\cdots,x_7,f_1,f_2,\cdots,f_8$ be a total of $15$ distinct elements from a finite field $\mathbb{F}_q, q\geq 15$, and let us set $L=36$, i.e., the message consists of $L=36$ i.i.d. symbols from $\mathbb{F}_q$ and is represented as  $\mathbf{W}^{(t)}=[W_1^{(t)},W_2^{(t)},\cdots,W_{36}^{(t)}]$. The Cauchy matrix generated by $x_1,x_2,\cdots,x_7,f_1,f_2,\cdots,f_8$ is defined as 
\begin{align}
\mathbf{C}=
\begin{bmatrix}
\frac{1}{x_1-f_1}&\frac{1}{x_1-f_2}&\cdots&\frac{1}{x_1-f_8}\\
\frac{1}{x_2-f_1}&\frac{1}{x_2-f_2}&\cdots&\frac{1}{x_2-f_8}\\
\vdots&\vdots&\cdots&\vdots\\
\frac{1}{x_7-f_1}&\frac{1}{x_7-f_2}&\cdots&\frac{1}{x_7-f_8}
\end{bmatrix}.
\end{align}
For all $t\in\mathbb{N}$, let $Z^{(t)}_1,Z^{(t)}_2,\cdots,Z^{(t)}_{48}$ be symbols from the finite field $\mathbb{F}_q$. In particular, for $t=0$, $Z^{(0)}_1,Z^{(0)}_2,\cdots,Z^{(0)}_{48}$ are uniformly i.i.d. over $\mathbb{F}_q$, independent of the (initial) message $\mathbf{W}^{(0)}$. For all $t\in\mathbb{N}$, let us define
\begin{align}
{\mathbf{M}}_1^{(t)}=\left[
\begin{NiceArray}{cccccc}
W^{(t)}_1 & W^{(t)}_2 & W^{(t)}_3 & W^{(t)}_4 & W^{(t)}_5 & W^{(t)}_6 \\
W^{(t)}_7 & W^{(t)}_8 & W^{(t)}_9 & W^{(t)}_{10} & W^{(t)}_{11} & W^{(t)}_{12}\\
W^{(t)}_{13} & W^{(t)}_{14} & W^{(t)}_{15} & W^{(t)}_{16} & W^{(t)}_{17} & W^{(t)}_{18}\\
W^{(t)}_{19} & W^{(t)}_{20} & W^{(t)}_{21} & W^{(t)}_{22} & W^{(t)}_{23} & W^{(t)}_{24}\\
W^{(t)}_{25} & W^{(t)}_{26} & W^{(t)}_{27} & W^{(t)}_{28} & W^{(t)}_{29} & W^{(t)}_{30}\\
\Block[transparent, fill=Plum!20,rounded-corners]{1-6}{} W^{(t)}_{31} & W^{(t)}_{32} & W^{(t)}_{33} & W^{(t)}_{34} & W^{(t)}_{35}  & W^{(t)}_{36}\\
\Block[transparent, fill=Cyan!20,rounded-corners]{1-6}{}
{Z^{(t)}_1}  & {Z^{(t)}_2}  & Z^{(t)}_{3} & Z^{(t)}_{4} & Z^{(t)}_{5} & {Z^{(t)}_6}\\
\Block[transparent, fill=ForestGreen!20,rounded-corners]{1-6}{}
{Z^{(t)}_7}  & {Z^{(t)}_8}  & Z^{(t)}_{9} & Z^{(t)}_{10} & Z^{(t)}_{11}  & {Z^{(t)}_{12}}
\end{NiceArray}\right],
\end{align}

\begin{align}
    \mathbf{M}_2^{(t)}=[{\mathbf{M}}_1^{'(t)}~{\mathbf{M}}_2^{'(t)}~{\mathbf{M}}_3^{'(t)}],
\end{align}

where
\begin{align}
{\mathbf{M}}_1^{'(t)}=\left[
\begin{NiceArray}{cc|c|ccc}
\Block[transparent, fill=Plum!20,rounded-corners]{3-2}{} W^{(t)}_{31} & W^{(t)}_{32}
& \Block[transparent, fill=Yellow!60,rounded-corners]{2-1}{}{Z^{(t)}_{13}} & \Block[transparent, fill=Orange!60,rounded-corners]{1-3}{}{Z^{(t)}_{15}} & {Z^{(t)}_{16}} &{Z^{(t)}_{18}} \\ 
W^{(t)}_{33} & W^{(t)}_{34}  & {Z^{(t)}_{14}}  & {Z^{(t)}_{19}} & {Z^{(t)}_{20}} & {Z^{(t)}_{21}}\\
W^{(t)}_{35} & W^{(t)}_{36}  & {Z^{(t)}_{17}} & {Z^{(t)}_{22}} & {Z^{(t)}_{23}} & {Z^{(t)}_{24}}\\
\Block[transparent, fill=Yellow!60,rounded-corners]{1-2}{}{Z^{(t)}_{13}} & {Z^{(t)}_{14}} & \Block[transparent, fill=Orange!60,rounded-corners]{1-1}{} {Z^{(t)}_{18}}  & 0 & 0 & 0\\
\Block[transparent, fill=Orange!60,rounded-corners]{1-2}{}{Z^{(t)}_{15}} & {Z^{(t)}_{16}} & 0 & 0 & 0 & 0\\
\end{NiceArray}\right],
\end{align}

\begin{align}
{\mathbf{M}}_2^{'(t)}=\left[
\begin{NiceArray}{cc|c|ccc}
\Block[transparent, fill=Cyan!20,rounded-corners]{3-2}{} Z^{(t)}_{1} & Z^{(t)}_{2}
& \Block[transparent, fill=Yellow!40,rounded-corners]{2-1}{}{Z^{(t)}_{25}} & \Block[transparent, fill=Orange!40,rounded-corners]{1-3}{}{Z^{(t)}_{27}} & {Z^{(t)}_{28}} &{Z^{(t)}_{30}} \\ 
Z^{(t)}_{3} & Z^{(t)}_{4}  & {Z^{(t)}_{26}}  & {Z^{(t)}_{31}} & {Z^{(t)}_{32}} & {Z^{(t)}_{33}}\\
Z^{(t)}_{5} & Z^{(t)}_{6}  & {Z^{(t)}_{29}} & {Z^{(t)}_{34}} & {Z^{(t)}_{35}} & {Z^{(t)}_{36}}\\
\Block[transparent, fill=Yellow!40,rounded-corners]{1-2}{}{Z^{(t)}_{25}} & {Z^{(t)}_{26}} & \Block[transparent, fill=Orange!40,rounded-corners]{1-1}{} {Z^{(t)}_{30}}  & 0 & 0 & 0\\
\Block[transparent, fill=Orange!40,rounded-corners]{1-2}{}{Z^{(t)}_{27}} & {Z^{(t)}_{28}} & 0 & 0 & 0 & 0\\
\end{NiceArray}\right],
\end{align}

\begin{align}
{\mathbf{M}}_3^{'(t)}=\left[
\begin{NiceArray}{cc|c|ccc}
\Block[transparent, fill=ForestGreen!20,rounded-corners]{3-2}{} Z^{(t)}_{7} & Z^{(t)}_{8}
& \Block[transparent, fill=Yellow!40,rounded-corners]{2-1}{}{Z^{(t)}_{37}} & \Block[transparent, fill=Orange!40,rounded-corners]{1-3}{}{Z^{(t)}_{39}} & {Z^{(t)}_{40}} &{Z^{(t)}_{42}} \\ 
Z^{(t)}_{9} & Z^{(t)}_{10}  & {Z^{(t)}_{38}}  & {Z^{(t)}_{43}} & {Z^{(t)}_{44}} & {Z^{(t)}_{45}}\\
Z^{(t)}_{11} & Z^{(t)}_{12}  & {Z^{(t)}_{41}} & {Z^{(t)}_{46}} & {Z^{(t)}_{47}} & {Z^{(t)}_{48}}\\
\Block[transparent, fill=Yellow!40,rounded-corners]{1-2}{}
{Z^{(t)}_{37}} & {Z^{(t)}_{38}} & \Block[transparent, fill=Orange!40,rounded-corners]{1-1}{} {Z^{(t)}_{42}}  & 0 & 0 & 0\\
\Block[transparent, fill=Orange!40,rounded-corners]{1-2}{}
{Z^{(t)}_{39}} & {Z^{(t)}_{40}} & 0 & 0 & 0 & 0\\
\end{NiceArray}\right].
\end{align}

At any time slot $t\in\mathbb{N}$, the storage at Server $n$, denoted as $\mathbf{S}_n^{(t)}$, is constructed as follows
\begin{align}\label{eq:nstor0}
\begin{array}{ll}
    \mathbf{S}_{1,n}^{(t)} = \mathbf{C}(n,:)\mathbf{M}_1^{(t)}, & ~\forall n\in [N], \\
    \mathbf{S}_{2,n}^{(t)} = \mathbf{C}(n,[N-S])\mathbf{M}_2^{(t)}, & ~\forall n\in [S+1:N],\\
    \mathbf{S}_{n}^{(t)}=\mathbf{S}_{1,n}^{(t)}, & ~\forall n\in[S],\\
    \mathbf{S}_{n}^{(t)}=(\mathbf{S}_{1,n}^{(t)},\mathbf{S}_{2,n}^{(t)}), & ~\forall n\in[S+1:N].
\end{array}
\end{align}

Note that for all $n \in [S]$, $H(\mathbf{S}_n^{(t)})/L=\frac{6}{36}=\frac{1}{6}$, i.e., the storage constraints are satisfied. Besides, for all $n \in [S+1:N]$, $H(\mathbf{S}_n^{(t)})/L=\frac{6+18}{36}=\frac{2}{3}$.

\subsubsection{Execution of the Read Operation}
Assume that the user wishes to retrieve the message $\mathbf{W}^{(t)}$ at time slot $t$ and $\mathcal{D}^{(t)}=\{3\}$, i.e., server $3$ drops out at this point. To this end, the user downloads the following symbols from available servers.
\begin{align}
\begin{array}{ll}
    \mathbf{A}_{1,n}^{(t)}=\mathbf{S}_{1,n}^{(t)}([6]),& n\in[7]\setminus\{3\},\\
    \mathbf{A}_{2,n}^{(t)}=\mathbf{S}_{2,n}^{(t)}([3]\cup[7:9]),& n\in[4:7],\\
    \mathbf{A}_{n}^{(t)}=\mathbf{A}_{1,n}^{(t)},& n\in[2],\\
    \mathbf{A}_n^{(t)}=(\mathbf{A}_{1,n}^{(t)},\mathbf{A}_{2,n}^{(t)}),& n\in[4:7].
\end{array}
\end{align}
According to the construction of the storage, the download symbols $\mathbf{A}_{2,n}^{(t)}([3]), n\in [4:7]$ can be represented as follows
\begin{align}
&\left[
\mathbf{A}^{(t)}_{2,4}([3])~
\mathbf{A}^{(t)}_{2,5}([3])~
\mathbf{A}^{(t)}_{2,6}([3])~
\mathbf{A}^{(t)}_{2,7}([3])~
\right]^\mathtt{T}\notag\\
=&\mathbf{C}([4:7],[5])\mathbf{M}_2^{(t)}([3])\notag\\
=&\begin{bmatrix}
\frac{1}{x_4-f_1}&\frac{1}{x_4-f_2}&\cdots&\frac{1}{x_4-f_5}\\
\frac{1}{x_5-f_1}&\frac{1}{x_5-f_2}&\cdots&\frac{1}{x_5-f_5}\\
\frac{1}{x_6-f_1}&\frac{1}{x_6-f_2}&\cdots&\frac{1}{x_6-f_5}\\
\frac{1}{x_7-f_1}&\frac{1}{x_7-f_2}&\cdots&\frac{1}{x_7-f_5}
\end{bmatrix}
\left[ \begin{NiceArray}{ccc}
W^{(t)}_{31} & W^{(t)}_{32} & Z^{(t)}_{13} \\
W^{(t)}_{33} & W^{(t)}_{34} & Z^{(t)}_{14} \\
W^{(t)}_{35} & W^{(t)}_{36} & Z^{(t)}_{17} \\
Z^{(t)}_{13} & Z^{(t)}_{14} & Z^{(t)}_{18} \\
Z^{(t)}_{15} & Z^{(t)}_{16} & 0
\end{NiceArray}\right].
\end{align}
Following a successive interference cancellation decoding strategy, it is therefore possible to recover $Z^{(t)}_{13},Z^{(t)}_{14},Z^{(t)}_{17},Z^{(t)}_{18}$ and $W^{(t)}_{31},W^{(t)}_{32},W^{(t)}_{33},W^{(t)}_{34}, W^{(t)}_{35},W^{(t)}_{36},Z^{(t)}_{15},Z^{(t)}_{16}$ successively. Specifically, recall that the square Cauchy matrix is invertible, we can recover the following symbols by solving the corresponding linear system
\begin{align}
    &\left[Z^{(t)}_{13}~Z^{(t)}_{14}~Z^{(t)}_{17}~Z^{(t)}_{18}\right]^\mathtt{T}= \notag\\
    &\mathbf{C}([4:7],[4])^{-1}\left[
\mathbf{A}^{(t)}_{2,4}(3)~
\mathbf{A}^{(t)}_{2,5}(3)~
\mathbf{A}^{(t)}_{2,6}(3)~
\mathbf{A}^{(t)}_{2,7}(3)~
\right]^\mathtt{T}.
\end{align}
Then, subtracting the decoded symbols $Z^{(t)}_{13},Z^{(t)}_{14}$ from $\mathbf{A}^{(t)}_{2,n}([2]),n\in [4:7]$, the other symbols are thus resolvable.
\begin{align}
    \begin{bmatrix}
        W^{(t)}_{31} & W^{(t)}_{32} \\
        W^{(t)}_{33} & W^{(t)}_{34} \\
        W^{(t)}_{35} & W^{(t)}_{36} \\
        Z^{(t)}_{15} & Z^{(t)}_{16}
    \end{bmatrix}
    =\mathbf{C}([4:7],\{1,2,3,5\})^{-1}\times\notag\\
    \begin{bmatrix}
        \mathbf{A}^{(t)}_{2,4}(1)-\frac{1}{x_4-f_4}Z^{(t)}_{13} &\mathbf{A}^{(t)}_{2,4}(2)-\frac{1}{x_4-f_4}Z^{(t)}_{14}\\
       \mathbf{A}^{(t)}_{2,5}(1)-\frac{1}{x_5-f_4}Z^{(t)}_{13} &\mathbf{A}^{(t)}_{2,5}(2)-\frac{1}{x_5-f_4}Z^{(t)}_{14}\\
       \mathbf{A}^{(t)}_{2,6}(1)-\frac{1}{x_6-f_4}Z^{(t)}_{13} &\mathbf{A}^{(t)}_{2,6}(2)-\frac{1}{x_6-f_4}Z^{(t)}_{14}\\
       \mathbf{A}^{(t)}_{2,7}(1)-\frac{1}{x_7-f_4}Z^{(t)}_{13} &\mathbf{A}^{(t)}_{2,7}(2)-\frac{1}{x_7-f_4}Z^{(t)}_{14}
    \end{bmatrix}.
\end{align}

Similarly, we can decode $Z^{(t)}_{1},Z^{(t)}_{2},Z^{(t)}_{3},Z^{(t)}_{4},Z^{(t)}_{5},Z^{(t)}_{6}$ from $\mathbf{A}_{2,n}^{(t)}([4:6]), n\in [4:7]$, and then the message symbols $W^{(t)}_{1},W^{(t)}_{2},\cdots,W^{(t)}_{30}$ are recoverable from $\mathbf{A}^{(t)}_{1,n}, n\in[7]\setminus\{3\}$.
The total download cost is
\begin{align}
    C_r^{(t)}=\frac{2\times6+4\times(6+6)}{36}=\frac{5}{3}.
\end{align}
Recall the definition $\mathcal{D}_1^{(t)}=[S+1:N]\cap\mathcal{D}^{(t)}$ and $\mathcal{D}_2^{(t)}=[S]\cap\mathcal{D}^{(t)}$, for this setting we have $|\mathcal{D}_1^{(t)}|=1$ and $|\mathcal{D}_2^{(t)}|=0$. Therefore, our scheme achieves the optimal download cost as
$\frac{N-S-|\mathcal{D}^{(t)}_1|}{N-|\mathcal{D}^{(t)}_1|-R_r+1}-\frac{(S-|\mathcal{D}^{(t)}_2|)(R_r-S-1)}{K_c(N-|\mathcal{D}^{(t)}_1|-R_r+1)}=2-\frac{1}{3}=\frac{5}{3}$.

\subsubsection{Execution of the Update Operation}
Assume that the user wishes to update the message with the local additive increment $\mathbf{\Delta}^{(t)}=[\Delta^{(t)}_1,\Delta^{(t)}_2,\cdots,\Delta^{(t)}_{36}]$ at time slot $t$. Suppose $\mathcal{D}^{(t)}=\{1\}$ and $X^{(t)}=1$, i.e., server $1$ drops out and the user wishes to prevent any single server from learning any information about the increment. To this end, let us define
\begin{align}
\dot{\mathbf{M}}_1^{(t)}=\left[
\begin{NiceArray}{cccccc}
\Delta^{(t)}_1 & \Delta^{(t)}_2 & \Delta^{(t)}_3 & \Delta^{(t)}_4 & \Delta^{(t)}_5 & \Delta^{(t)}_6 \\
\Delta^{(t)}_7 & \Delta^{(t)}_8 & \Delta^{(t)}_9 & \Delta^{(t)}_{10} & \Delta^{(t)}_{11} & \Delta^{(t)}_{12}\\
\Delta^{(t)}_{13} & \Delta^{(t)}_{14} & \Delta^{(t)}_{15} & \Delta^{(t)}_{16} & \Delta^{(t)}_{17} & \Delta^{(t)}_{18}\\
\Delta^{(t)}_{19} & \Delta^{(t)}_{20} & \Delta^{(t)}_{21} & \Delta^{(t)}_{22} & \Delta^{(t)}_{23} & \Delta^{(t)}_{24}\\
\Delta^{(t)}_{25} & \Delta^{(t)}_{26} & \Delta^{(t)}_{27} & \Delta^{(t)}_{28} & \Delta^{(t)}_{29} & \Delta^{(t)}_{30}\\
\Block[transparent, fill=Plum!20,rounded-corners]{1-6}{} \Delta^{(t)}_{31} & \Delta^{(t)}_{32} & \Delta^{(t)}_{33} & \Delta^{(t)}_{34} & \Delta^{(t)}_{35}  & \Delta^{(t)}_{36}\\
\Block[transparent, fill=Cyan!20,rounded-corners]{1-6}{}
\dot{Z}^{(t)}_1 & \dot{Z}^{(t)}_2  & \dot{Z}^{(t)}_{3} & \dot{Z}^{(t)}_{4} & \dot{Z}^{(t)}_{5} & \dot{Z}^{(t)}_6\\
\Block[transparent, fill=ForestGreen!20,rounded-corners]{1-6}{}
{H^{(t)}_1}  & {H^{(t)}_2}  &{H^{(t)}_3}  & {H^{(t)}_4} &{H^{(t)}_5} & {H^{(t)}_6}
\end{NiceArray}\right],
\end{align}
\begin{align}
    \dot{\mathbf{M}}_2^{(t)}=[\dot{\mathbf{M}}_1^{'(t)}~\dot{\mathbf{M}}_2^{'(t)}~\dot{\mathbf{M}}_3^{'(t)}],
\end{align}
where
\begin{align}
\dot{\mathbf{M}}_1^{'(t)}=\left[
\begin{NiceArray}{cc|c|ccc}
\Block[transparent, fill=Plum!20,rounded-corners]{3-2}{} \Delta^{(t)}_{31} & \Delta^{(t)}_{32}
& \Block[transparent, fill=Yellow!60,rounded-corners]{2-1}{} \dot{Z}^{(t)}_{13} & \Block[transparent, fill=Orange!60,rounded-corners]{1-3}{} 0 & 0 & 0 \\ 
\Delta^{(t)}_{33} & \Delta^{(t)}_{34}  & \dot{Z}^{(t)}_{14}  &  0 & 0 & 0\\
\Delta^{(t)}_{35} & \Delta^{(t)}_{36}  & \dot{Z}^{(t)}_{17} &  0 & 0 & 0\\
\Block[transparent, fill=Yellow!60,rounded-corners]{1-2}{} \dot{Z}^{(t)}_{13} & \dot{Z}^{(t)}_{14} & \Block[transparent, fill=Orange!60,rounded-corners]{1-1}{}  0  & 0 & 0 & 0\\
\Block[transparent, fill=Orange!60,rounded-corners]{1-2}{} 0 & 0 & 0 & 0 & 0 & 0
\end{NiceArray}\right],
\end{align}
\begin{align}
\dot{\mathbf{M}}_2^{'(t)}=\left[
\begin{NiceArray}{cc|c|ccc}
\Block[transparent, fill=Cyan!20,rounded-corners]{3-2}{} \dot{Z}^{(t)}_1 & \dot{Z}^{(t)}_2
& \Block[transparent, fill=Yellow!40,rounded-corners]{2-1}{} \dot{Z}^{(t)}_{25} & \Block[transparent, fill=Orange!40,rounded-corners]{1-3}{} 0 & 0 & 0 \\ 
\dot{Z}^{(t)}_{3} & \dot{Z}^{(t)}_{4}  & \dot{Z}^{(t)}_{26}  &  0 & 0 & 0\\
\dot{Z}^{(t)}_{5} & \dot{Z}^{(t)}_6  & \dot{Z}^{(t)}_{29} &  0 & 0 & 0\\
\Block[transparent, fill=Yellow!40,rounded-corners]{1-2}{} \dot{Z}^{(t)}_{25} & \dot{Z}^{(t)}_{26} & \Block[transparent, fill=Orange!40,rounded-corners]{1-1}{} 0  & 0 & 0 & 0\\
\Block[transparent, fill=Orange!40,rounded-corners]{1-2}{} 0 & 0 & 0 & 0 & 0 & 0
\end{NiceArray}\right],
\end{align}
\begin{align}
\dot{\mathbf{M}}_3^{'(t)}=\left[
\begin{NiceArray}{cc|c|ccc}
\Block[transparent, fill=ForestGreen!20,rounded-corners]{3-2}{} {H^{(t)}_1}  & {H^{(t)}_2}
& \Block[transparent, fill=Yellow!20,rounded-corners]{2-1}{} \dot{Z}^{(t)}_{37} & \Block[transparent, fill=Orange!20,rounded-corners]{1-3}{} 0 & 0 & 0 \\ 
{H^{(t)}_3}  & {H^{(t)}_4}  & \dot{Z}^{(t)}_{38}  &  0 & 0 & 0\\
{H^{(t)}_5} & {H^{(t)}_6} & \dot{Z}^{(t)}_{41} &  0 & 0 & 0\\
\Block[transparent, fill=Yellow!20,rounded-corners]{1-2}{} \dot{Z}^{(t)}_{37} & \dot{Z}^{(t)}_{38} & \Block[transparent, fill=Orange!20,rounded-corners]{1-1}{} 0  & 0 & 0 & 0\\
\Block[transparent, fill=Orange!20,rounded-corners]{1-2}{} 0 & 0 & 0 & 0 & 0 & 0
\end{NiceArray}\right],
\end{align}
and $\dot{Z}_i^{(t)}$,$i\in\{1,2,\cdots,6,13,14,17,25,26,29,37,38,41\}$ are uniformly i.i.d. symbols from $\mathbb{F}_q$, independent of the increment $\mathbf{\Delta}^{(t)}$, used to guarantee the $X^{(t)}=1$-security, and $H_1^{(t)},H_2^{(t)},\cdots,H_6^{(t)}$ are constructed to guarantee that
\begin{align}
    \mathbf{C}(\mathcal{D}^{(t)},:)\dot{\mathbf{M}}_1^{(t)}=\mathbf{0}.
\end{align}
Recall that this is made possible by setting
\begin{align}
    H_n^{(t)}=-(x_1-f_8)\big(\sum_{i=1}^{6}\frac{1}{x_1-f_i}{\Delta}^{(t)}_{6(i-1)+n}+\frac{1}{x_1-f_7}Z_n^{(t)} \big)
\end{align}
for $n\in[6]$. The construction of the coded increment $\mathbf{Q}_n^{(t)},n\in[N]\setminus\mathcal{D}^{(t)}$ is
\begin{align}
\begin{array}{ll}
    \mathbf{Q}_{1,n}^{(t)}=\mathbf{C}(n,:)\dot{\mathbf{M}}_1^{(t)},& ~\forall n\in[2:7], \\
    \mathbf{Q}_{2,n}^{(t)} = \mathbf{C}(n,[N-S])\dot{\mathbf{M}}_2^{(t)}, & ~\forall n\in[3:7],\\
    \mathbf{Q}_{n}^{(t)}=\mathbf{Q}_{1,n}^{(t)}, & ~\forall n\in\{2\},\\
    \mathbf{Q}_{n}^{(t)}=(\mathbf{Q}_{1,n}^{(t)},\mathbf{Q}_{2,n}^{(t)}), & ~\forall n\in[3:7].\\
\end{array}
\end{align}
Upon receiving the coded increment, Server $n, n\in[N]\setminus\mathcal{D}^{(t)}$ updates its storage according to the following equation
\begin{align}
    \mathbf{S}_n^{(t+1)}=&\mathbf{S}^{(t)}_n+\mathbf{Q}^{(t)}_n.
\end{align}
Due to the fact that at time slot $t+1$, we have $\mathbf{M}_1^{(t+1)}=\mathbf{M}_1^{(t)}+\dot{\mathbf{M}}_1^{(t)}$ and $\mathbf{M}_2^{(t+1)}=\mathbf{M}_2^{(t)}+\dot{\mathbf{M}}_2^{(t)}$, the scheme updates the message correctly, while guaranteeing that the storage at dropout servers is left untouched. Besides, the update operation is feasible when 6 servers are available, which matches the update threshold $R_u^{(t)}=N-R_r+S+1+X^{(t)}=6$. The $X^{(t)}$-security guaranteed by the MDS$(7,1)$-coded noise symbols.
The total normalized upload cost of our scheme is 
\begin{align}
    C_u^{(t)}=\frac{6+5\times(6+9)}{36}=\frac{9}{4}.
\end{align}
Since $|\mathcal{D}_1^{(t)}|=0,|\mathcal{D}_2^{(t)}|=1,X^{(t)}=1$, we have $\frac{N-S-|\mathcal{D}^{(t)}_1|}{R_r-S-|\mathcal{D}^{(t)}_1|-X^{(t)}}-\frac{(S-|\mathcal{D}^{(t)}_2|)(N-R_r+X^{(t)})}{K_c(R_r-S-|\mathcal{D}^{(t)}_1|-X^{(t)})}=\frac{5}{2}-\frac{1}{4}=\frac{9}{4}$, i.e., our scheme achieves the optimal upload cost.

\subsection{General Scheme}
As discussed in Remark \ref{remark:homogeneous}, in this section, we only consider the setting where $S<K_c$. Recall that for this setting, we have $\Omega=R_r-S-1$. Define
\begin{align*}
        &\alpha_1=\max(K_c,N-\Omega), &&\gamma_1=L/\alpha_1, \\
        &G_1=\alpha_1-K_c+1, &&\alpha_i=\alpha_1-i+1, \\
        &\beta_i=\alpha_1+\Omega-i+1, &&\lambda_i=L/\alpha_i,~\forall i\in[G_1], \\
        &\gamma_i=L/(\alpha_i\alpha_{i-1}),&&\forall i\in[2:G_1],G_1\ge 2,  \\
        &L'=L/K_c, && \gamma'_1=L'/\alpha'_1, \\
        &G_2=N-R_r+1, &&\alpha'_i=N-S-\Omega-i+1, \\
        &\beta'_i=N-S-i+1,
        &&\lambda'_i=L'/\alpha'_i,~\forall i\in[G_2],\\
        &\gamma'_i=L'/(\alpha'_i\alpha'_{i-1}), && \forall i\in[2:G_2],G_2\ge 2,\\
        &\lambda_0=\lambda'_0=0, &&{P=K_c-S-1}.
\end{align*}
Note that $\lambda_i=\sum_{j=1}^i \gamma_j,\forall i\in [G_1]$ and $\lambda'_i=\sum_{j=1}^i \gamma'_j, \forall i \in [G_2]$. And the size of the message $L$ is set as the smallest integer satisfying $\text{lcm}((\alpha_{i})_{i\in[G_1]})\mid L$ and $\text{lcm}((\alpha'_{i})_{i\in[G_2]})\mid L'$.

\subsubsection{Construction of the Storage}
Let the message at any time slot $t$ consist of $L$ symbols from the finite field $\mathbb{F}_q$, defined as $\mathbf{W}^{(t)}=[W_1^{(t)},W_2^{(t)},\cdots,W_L^{(t)}]$. Let $x_1,x_2,\cdots,x_N,f_1,f_2,\cdots,f_{\beta_1}$ be a total of $N+\beta_1$ distinct elements from the finite field $\mathbb{F}_q,q\geq N+\beta_1$. 
{And the Cauchy matrix generated by these constants is defined as follows}
\begin{align}\label{eq:ccmat}
\mathbf{C}=
\begin{bmatrix}
\frac{1}{x_1-f_1}&\frac{1}{x_1-f_2}&\cdots&\frac{1}{x_1-f_{\beta_1}}\\
\frac{1}{x_2-f_1}&\frac{1}{x_2-f_2}&\cdots&\frac{1}{x_2-f_{\beta_1}}\\
\vdots&\vdots&\cdots&\vdots\\
\frac{1}{x_N-f_1}&\frac{1}{x_N-f_2}&\cdots&\frac{1}{x_N-f_{\beta_1}}
\end{bmatrix}.
\end{align}
We then define matrices $\mathbf{M}_1^{(t)},\mathbf{M}_2^{(t)}$ as the ones generated by $\FuncSty{PSCGen}(\mathbf{W}^{(t)},(\mathbf{Z}_i)_{i\in [G_1]},(\mathbf{Z}_{i,k})_{\forall i\in[P],j\in[G_2]})$, shown in Algorithm \ref{alg:recode}. {And $\FuncSty{SCGen}(\mathbf{W}, (\mathbf{Z}_i)_{i\in[G]},(\alpha_i,\gamma_i,\beta_i)_{i\in[G]})$, shown in Algorithm \ref{alg:scgen}, is used as its fundamental building block.}

\begin{algorithm}
\DontPrintSemicolon
\caption{Generation of Staircase Structure}\label{alg:scgen}
\KwIn{$\mathbf{W}\in\mathbb{F}_q^{L},$ $\mathbf{Z}_i\in\mathbb{F}_q^{\Omega\times\gamma_i},(\alpha_i,\gamma_i,\beta_i)_{i\in[G]}$}
\KwOut{$\mathbf{M}=\left[\mathbf{M}_1,\mathbf{M}_2,\cdots,\mathbf{M}_G\right]$}
\newcommand{\nonl}{\renewcommand{\nl}{\let\nl\oldnl}}
\SetAlgoNoEnd
\SetKwFunction{FMain}{SCGen}
    \SetKwProg{Fn}{Function}{:}{}
    \Fn{\FMain{$\mathbf{W}, (\mathbf{Z}_i)_{i\in[G]},(\alpha_i,\gamma_i,\beta_i)_{i\in[G]}$}}{
        \ForEach{$i\in[G]$}{
            \eIf{ $i=1$}{
                $\mathbf{M}_1([\alpha_1],:)\gets\FuncSty{Reshape}(\mathbf{W},\alpha_1,\gamma_1)$\;
                $\mathbf{M}_1([\alpha_1+1:N],:)\gets\mathbf{Z}_1$\;
            }{
                $\mathbf{D}_{i-1}\gets\FuncSty{Reshape}([\mathbf{M}_1(R_r+i-1,:),\cdots,\allowbreak\mathbf{M}_j(R_r+i-j,:),\cdots,\mathbf{M}_{i-1}(R_r+1,:)],\alpha_i,\gamma_i)$\;
                $\mathbf{M}_i([\alpha_i],:)\gets\mathbf{D}_{i-1}$\;
                $\mathbf{M}_i([\alpha_i+1:\beta_i],:)\gets\mathbf{Z}_{i}$\;
                $\mathbf{M}_i([\beta_i+1:N],:)\gets\mathbf{0}_{(N-\beta_i)\times \gamma_i}$\;
            }
        } 
        \textbf{return} $\mathbf{M}=\left[\mathbf{M}_1,\mathbf{M}_2,\cdots,\mathbf{M}_G\right]$\;
    }
    \textbf{End Function}
\end{algorithm}
\begin{algorithm}
\DontPrintSemicolon
\SetAlgoNoEnd
\caption{Generation of Partially Constrained Storage Structure}\label{alg:recode}
\KwIn{$\mathbf{W},(\mathbf{Z}_i)_{i\in [G_1]},(\mathbf{Z}_{i,j})_{i\in[P],j\in[G_2]})$}
\KwOut{$\mathbf{M}_1,\mathbf{M}_2$}
\SetKwFunction{FMain}{PSCGen}
    \SetKwProg{Fn}{Function}{:}{}
    \Fn{\FMain{$\mathbf{W},(\mathbf{Z}_i)_{i\in [G_1]},(\mathbf{Z}_{i,j})_{i\in[P],j\in[G_2]}$}}
    {   
        $\mathbf{M}_1=\left[\mathbf{M}_{1,1},\mathbf{M}_{1,2},\cdots,\mathbf{M}_{1,G_1}\right]\gets\FuncSty{SCGen}(\mathbf{W},(\mathbf{Z}_i)_{i\in [G_1]},(\alpha_i,\gamma_i,\beta_i)_{i\in[G_1]})$\;        
        \ForEach(\Comment{Re-encoding mechanism}){$i\in [P]$}
        {
        $\mathbf{W}'_{i}\gets [\mathbf{M}_{1,1}(R_r+G_1-1+i,:),$$\cdots,\mathbf{M}_{1,j}(R_r+G_1-1+i-j,:),$$\cdots,\mathbf{M}_{1,G_1}(R_r-1+i,:)]$\; 
        $\mathbf{M}'_{i}=\left[\mathbf{M}'_{i,1},\mathbf{M}'_{i,2},\cdots,\mathbf{M}'_{i,G_2}\right]\gets\FuncSty{SCGen}(\mathbf{W}_{i}',(\mathbf{Z}_{i,j})_{j\in[G_2]},(\alpha'_j,\gamma'_j,\beta'_j)_{j\in[G_2]})$\;    }
        $\mathbf{M}_2=\left[\mathbf{M}'_{1}~\mathbf{M}'_{2}~\cdots~\mathbf{M}'_{P}\right]$ \;
    \textbf{return} $\mathbf{M}_1,\mathbf{M}_2$\;
    }
    \textbf{End Function}
\end{algorithm}\vspace{-0.25cm}

Consequently, at any time slot $t\in\mathbb{N}$, the storage at Server $n$ is defined by 
\begin{align}
    &\mathbf{S}_{1,n}^{(t)} = \mathbf{C}(n,:)\mathbf{M}_1^{(t)}, && ~\forall n\in [N], \label{storage-1}\\
    &\mathbf{S}_{2,n}^{(t)} = \mathbf{C}(n,[N-S])\mathbf{M}_2^{(t)}, && ~\forall n\in [S+1:N],\\
    &\mathbf{S}_{n}^{(t)}=\mathbf{S}_{1,n}^{(t)}, && ~\forall n\in[S],\\
    &\mathbf{S}_{n}^{(t)}=(\mathbf{S}_{1,n}^{(t)},\mathbf{S}_{2,n}^{(t)}), && ~\forall n\in[S+1:N].
\end{align}
It can be verified that for all $n \in [S]$, we have $H(\mathbf{S}_n^{(t)})/L=\frac{1}{K_c}$, i.e., the storage constraints are satisfied. Besides, for all $n \in [S+1:N]$, $H(\mathbf{S}_n^{(t)})/L=\frac{1}{K_c}+\frac{P\lambda'_{G_2}}{L}=1-\frac{S}{K_c}$.

\subsubsection{Execution of the Read Operation}
Suppose that the user wishes to execute the read operation at time slot $t$, let us consider the following two cases.

\paragraph {When $|\mathcal{D}^{(t)}|\leq N-\Omega-K_c$}
Define $J^{(t)}=|\mathcal{D}^{(t)}|+1$, the download symbols from server $n$ are
\begin{align}
    \mathbf{A}_n^{(t)}=\mathbf{S}_n^{(t)}([\lambda_{J^{(t)}}]), n\in[N]\backslash\mathcal{D}^{(t)}.
\end{align}
 
Subsequently, the message can be decoded via a \textit{successive interference cancellation strategy}. 
Define the available server as ${\overline{\mathcal{D}}^{(t)}=[N]\setminus\mathcal{D}^{(t)}}$, recalling the construction of the storage, the downloaded symbols can be written in the following matrix form
\begin{align}
\begin{bmatrix}
\mathbf{A}^{(t)}_{\overline{\mathcal{D}}^{(t)}(1)}\\
\mathbf{A}^{(t)}_{\overline{\mathcal{D}}^{(t)}(2)}\\
\vdots\\
\mathbf{A}^{(t)}_{\overline{\mathcal{D}}^{(t)}(|\overline{\mathcal{D}}^{(t)}|)}
\end{bmatrix}
=&\mathbf{C}(\overline{\mathcal{D}}^{(t)},:)\mathbf{M}_1^{(t)}(:,[\lambda_{J^{(t)}}])\\
=&\mathbf{C}(\overline{\mathcal{D}}^{(t)},:)\left[\mathbf{M}^{(t)}_{1,1},\mathbf{M}^{(t)}_{1,2},\cdots,\mathbf{M}^{(t)}_{1,J^{(t)}}\right].
\end{align}
Since for all $i\in[J^{(t)}]$, $\mathbf{M}^{(t)}_{1,i}([\beta_i+1:N],:)$ are zeros according to Algorithm \ref{alg:scgen}, for all $i\in[J^{(t)}]$, we have
\begin{align}
\begin{bmatrix}
\mathbf{A}^{(t)}_{\overline{\mathcal{D}}^{(t)}(1)}([\lambda_{i-1}+1:\lambda_{i}])\\
\mathbf{A}^{(t)}_{\overline{\mathcal{D}}^{(t)}(2)}([\lambda_{i-1}+1:\lambda_{i}])\\
\vdots\\
\mathbf{A}^{(t)}_{\overline{\mathcal{D}}^{(t)}(|\overline{\mathcal{D}}^{(t)}|)}([\lambda_{i-1}+1:\lambda_{i}])
\end{bmatrix}=
    \mathbf{C}(\overline{\mathcal{D}}^{(t)},[\beta_{i}])\mathbf{M}^{(t)}_{1,i}([\beta_{i}],:).
\end{align}
Due to the invertibility of square Cauchy matrices, $\mathbf{M}^{(t)}_{1,J^{(t)}}([\beta_{J^{(t)}}],:)$ is resolvable by
\begin{align}\label{eq:decMJ}
\mathbf{M}^{(t)}_{1,J^{(t)}}([\beta_{J^{(t)}}],:)&=
\mathbf{C}^{-1}(\overline{\mathcal{D}}^{(t)},[\beta_{J^{(t)}}])\times\begin{bmatrix}
\mathbf{A}^{(t)}_{\overline{\mathcal{D}}^{(t)}(1)}([\lambda_{J^{(t)}-1}+1:\lambda_{J^{(t)}}])\\
\mathbf{A}^{(t)}_{\overline{\mathcal{D}}^{(t)}(2)}([\lambda_{J^{(t)}-1}+1:\lambda_{J^{(t)}}])\\
\vdots\\
\mathbf{A}^{(t)}_{\overline{\mathcal{D}}^{(t)}(|\overline{\mathcal{D}}^{(t)}|)}([\lambda_{J^{(t)}-1}+1:\lambda_{J^{(t)}}])
\end{bmatrix}.
\end{align}
Due to the replication structure according to lines 7-8 of Algorithm \ref{alg:scgen}, if $(\mathbf{M}^{(t)}_{1,i'})_{i'\in[i+1:J^{(t)}]}$ is decoded, then the corresponding elements of $\mathbf{M}^{(t)}_{1,i}([R_r+1:R_r+J^{(t)}-i],:)$ are also recoverable, and their contribution to the downloaded symbols can be subtracted. Recall that $|[\beta_i]\setminus[R_r+1:R_r+J^{(t)}-i]|=\beta_i-(J^{(t)}-i)=|\overline{\mathcal{D}}^{(t)}|$, therefore $\mathbf{M}^{(t)}_{1,i}([\beta_i]\setminus[R_r+1:R_r+J^{(t)}-i],:)$ can be decoded by subtracting the recovered symbols and inverting the square Cauchy matrix on the LHS of \eqref{eq:decMi} as follows
\begin{align}\label{eq:decMi}
&\mathbf{C}(\overline{\mathcal{D}}^{(t)},[\beta_i]\setminus[R_r+1:R_r+J^{(t)}-i])\times\mathbf{M}^{(t)}_{1,i}([\beta_i]\setminus[R_r+1:R_r+J^{(t)}-i],:)=\notag\\
&\begin{bmatrix}
\mathbf{A}^{(t)}_{\overline{\mathcal{D}}^{(t)}(1)}([\lambda_{i-1}+1:\lambda_{i}])\notag\\
\mathbf{A}^{(t)}_{\overline{\mathcal{D}}^{(t)}(2)}([\lambda_{i-1}+1:\lambda_{i}])\\
\vdots\\
\mathbf{A}^{(t)}_{\overline{\mathcal{D}}^{(t)}(|\overline{\mathcal{D}}^{(t)}|)}([\lambda_{i-1}+1:\lambda_{i}])
\end{bmatrix}-\mathbf{C}(\overline{\mathcal{D}}^{(t)},[R_r+1:R_r+J^{(t)}-i])\times\mathbf{M}^{(t)}_{1,i}([R_r+1:R_r+J^{(t)}-i],:).
\end{align}
In other words, $\mathbf{M}_{1,i}^{(t)}$ is recovered at this point. Now it is evident to see that $\mathbf{M}_{1,J^{(t)}}^{(t)}, \mathbf{M}_{1,J^{(t)}-1}^{(t)},\cdots, \mathbf{M}_{1,1}^{(t)}$ are recoverable by induction on $i$, from which the message $\mathbf{W}^{(t)}$ is decodable.

Since $\alpha_1=\max(K_c, N-\Omega)$, for this setting, we have $N-\Omega-K_c\geq|\mathcal{D}^{(t)}|>0$, i.e., $\alpha_1=N-\Omega$. Therefore, $\lambda_{J^{(t)}}=\frac{L}{\alpha_J^{(t)}}=\frac{L}{N-\Omega-J^{(t)}+1}=\frac{L}{N-|\mathcal{D}^{(t)}|-R_r+S+1}$, and the normalized download cost can be calculated as
\begin{align}
    C^{(t)}_r=\frac{(N-|\mathcal{D}^{(t)}|)\lambda_{J^{(t)}}}{L}=\frac{N-|\mathcal{D}^{(t)}|}{N-|\mathcal{D}^{(t)}|-R_r+S+1}.
\end{align}
To show the optimality, consider the bound \eqref{eq:defD-1}, averaging over all possible choices of $\mathcal{X}\subset[N]\setminus\mathcal{D}^{(t)},|\mathcal{X}|=R_r-S-1$, we have
\begin{align}
    \binom{N-|\mathcal{D}|-1}{|\mathcal{X}|}\sum_{n\in[N]\setminus\mathcal{D}}D_n&\geq\binom{N-|\mathcal{D}|}{|\mathcal{X}|}.
\end{align}
Expanding the binomial coefficients, we have
\begin{align}
    \frac{(N-|\mathcal{D}|-1)!}{|\mathcal{X}|!(N-|\mathcal{D}|-|\mathcal{X}|-1)!}\sum_{n\in[N]\setminus\mathcal{D}}D_n&\geq\frac{(N-|\mathcal{D}|)!}{|\mathcal{X}|!(N-|\mathcal{D}|-|\mathcal{X}|)!}\\
    \sum_{n\in[N]\setminus\mathcal{D}}D_n &\ge\frac{N-|\mathcal{D}^{(t)}|}{N-|\mathcal{D}^{(t)}|-|\mathcal{X}|} \\
    &=\frac{N-|\mathcal{D}^{(t)}|}{N-|\mathcal{D}^{(t)}|-R_r+S+1}.
\end{align}
Thus, we have $
    C^{(t)}_r\geq\min_{(D_1, \cdots, D_N)\in\mathcal{D}}\sum_{n\in[N]\setminus\mathcal{D}^{(t)}}D_n 
    \geq\frac{N-|\mathcal{D}^{(t)}|}{N-|\mathcal{D}^{(t)}|-R_r+S+1}$,
i.e., our scheme achieves the lower bound on $C_r^{(t)}$.
\paragraph {When $|\mathcal{D}^{(t)}|> N-\Omega-K_c$}
Define $ J_1^{(t)}=K_c+\Omega-N+|\mathcal{D}^{(t)}|,J_2^{(t)}=|\mathcal{D}_1^{(t)}|+1$ and 
$\mu_{J_1^{(t)},J_2^{(t)}}=\bigcup_{i=0}^{J_1^{(t)}-1}[i\times\lambda'_{G_2}+1:i\times\lambda'_{G_2}+\lambda'_{J_2^{(t)}}]$.
The download symbols from server $n$ are
\begin{align}
    &\mathbf{A}_{1,n}^{(t)}=\mathbf{S}_{1,n}^{(t)}([\lambda_{G_1}]),&& n\in[N]\setminus\mathcal{D}^{(t)},\\
    &\mathbf{A}_{2,n}^{(t)}=\mathbf{S}_{2,n}^{(t)}(\mu_{J_1^{(t)},J_2^{(t)}}), &&n\in[S+1:N]\setminus\mathcal{D}^{(t)},\\
    &\mathbf{A}_n^{(t)}=\mathbf{A}_{1,n}^{(t)}, &&n\in[S]\setminus\mathcal{D}^{(t)},\\
    &\mathbf{A}_n^{(t)}=(\mathbf{A}_{1,n}^{(t)},\mathbf{A}_{2,n}^{(t)}), &&n\in[S+1:N]\setminus\mathcal{D}^{(t)}.
\end{align}
According to line 3-5 of Algorithm \ref{alg:recode}, the symbols $\mathbf{M}^{(t)}_{1,1}(R_r+G_1:R_r+G_1-1+J_1^{(t)},:),\cdots,\mathbf{M}_{1,j}^{(t)}(R_r+G_1-j:R_r+G_1-j-1+J_1^{(t)},:),\cdots,\mathbf{M}^{(t)}_{1,G_1}(R_r:R_r-1+J_1^{(t)},:)$ can be recovered from $\mathbf{A}_{2,n}^{(t)},n\in[S+1:N]\setminus\mathcal{D}^{(t)}$. by following a similar \textit{successive interference cancellation strategy} as described in the preceding subsection. Then, $\mathbf{W}^{(t)}$ can be recovered from $\overline{\mathbf{A}}_{1,n}^{(t)},n\in[N]\setminus\mathcal{D}^{(t)}$ using the same strategy, where for all $n\in[N]\setminus\mathcal{D}^{(t)}$, $\overline{\mathbf{A}}_{1,n}^{(t)}
    =\mathbf{A}_{1,n}^{(t)}-(\sum_{j\in[G_1]}\mathbf{C}(n,R_r+G_1+1-j:R_r+G_1+1-j+J_1^{(t)})\times\mathbf{M}^{(t)}_{1,j}(R_r+G_1+1-j:R_r+G_1+1-j+J_1^{(t)},:))$. 
    For this setting, $\lambda'_{{J}_2^{(t)}}=\frac{L'}{\alpha'_{J_2^{(t)}}}=\frac{L}{K_c(N-S-\Omega-J_2^{(t)}+1)}=\frac{L}{K_c(N-S-\Omega-|\mathcal{D}_1^{(t)}|)}=\frac{L}{K_c(N-|\mathcal{D}_1^{(t)}|-R_r+1)}$. Therefore, the normalized download cost is
\begin{align}
    &C^{(t)}_r=\frac{S-|\mathcal{D}_2^{(t)}|}{K_c}+(N-S-|\mathcal{D}_1^{(t)}|)(\frac{1}{K_c}+\frac{J_1^{(t)}\lambda'_{{J}_2^{(t)}}}{L})\\
    &=\frac{S-|\mathcal{D}_2^{(t)}|}{K_c}+\frac{(N-S-|\mathcal{D}_1^{(t)}|)(K_c+|\mathcal{D}^{(t)}_2|-S)}{K_c(N-|\mathcal{D}_1^{(t)}|-R_r+1)}\\
    &=\frac{N-S-|\mathcal{D}^{(t)}_1|}{N-|\mathcal{D}^{(t)}_1|-R_r+1}-\frac{(S-|\mathcal{D}^{(t)}_2|)(R_r-S-1)}{K_c(N-|\mathcal{D}^{(t)}_1|-R_r+1)}.
\end{align}
To show the optimality, consider the bound in \eqref{eq:defD-1} over all possible choices of $\mathcal{X}\subset[N]\setminus\mathcal{D}^{(t)}\setminus[S],|\mathcal{X}|=R_r-S-1$, we have
\begin{align}
    &\binom{N-S-|\mathcal{D}^{(t)}_1|-1}{|\mathcal{X}|}\sum_{n\in[S+1:N]\setminus\mathcal{D}^{(t)}}D_n+\binom{N-S-|\mathcal{D}^{(t)}_1|}{|\mathcal{X}|}\sum_{n\in[S]\setminus\mathcal{D}^{(t)}}D_n\geq\binom{N-S-|\mathcal{D}^{(t)}_1|}{|\mathcal{X}|}, \\
    &\Rightarrow \sum_{n\in[S+1:N]\setminus\mathcal{D}^{(t)}}D_n+\frac{N-S-|\mathcal{D}^{(t)}_1|}{N-S-|\mathcal{D}^{(t)}_1|-|\mathcal{X}|}\sum_{n\in[S]\setminus\mathcal{D}^{(t)}}D_n\geq\frac{N-S-|\mathcal{D}^{(t)}_1|}{N-S-|\mathcal{D}^{(t)}_1|-|\mathcal{X}|}
    \label{lp-d1}
\end{align}
by expanding binomial coefficients. In addition, according to \eqref{eq:defD}, we have
\begin{align}
    -\frac{|\mathcal{X}|}{N-S-|\mathcal{D}^{(t)}_1|-|\mathcal{X}|}\sum_{n\in[S]\setminus\mathcal{D}^{(t)}}D_n\geq-\frac{(S-|\mathcal{D}^{(t)}_2|)|\mathcal{X}|}{K_c(N-S-|\mathcal{D}^{(t)}_1|-|\mathcal{X}|)}\label{lp-d2}.
\end{align}
Combining the above bounds, we have
\begin{align}
    &C^{(t)}_r\ge \min_{(D_1, \cdots, D_N)\in\mathcal{D}}\sum_{n\in[N]\setminus\mathcal{D}^{(t)}}D_n\\
     \ge&\frac{N-S-|\mathcal{D}^{(t)}_1|}{N-S-|\mathcal{D}^{(t)}_1|-|\mathcal{X}|}-\frac{(S-|\mathcal{D}^{(t)}_2|)|\mathcal{X|}}{K_c(N-S-|\mathcal{D}^{(t)}_1|-|\mathcal{X}|)}\\
    =&\frac{N-S-|\mathcal{D}^{(t)}_1|}{N-|\mathcal{D}^{(t)}_1|-R_r+1}-\frac{(S-|\mathcal{D}^{(t)}_2|)(R_r-S+1)}{K_c(N-|\mathcal{D}^{(t)}_1|-R_r+1)},
\end{align}
i.e., the normalized download cost of our scheme is optimal.

\subsubsection{Execution of the Update Operation}
Assume that the user wishes to update the message at time slot $t$ with the increment $\mathbf{\Delta}^{(t)}=[\Delta_1^{(t)},\Delta_2^{(t)},\cdots,\Delta_L^{(t)}]$.

\paragraph {When $|\mathcal{D}^{(t)}|+X^{(t)}\le R_r-K_c$}
Define $G_1^{(t)}=\alpha_1+1-R_r+|\mathcal{D}^{(t)}|+X^{(t)}$.
Let $\dot{\mathbf{M}}_1^{(t)},\dot{\mathbf{M}}_2^{(t)}=\FuncSty{PSCGen}(\mathbf{\Delta}^{(t)},(\dot{\mathbf{Z}}_i^{(t)})_{i\in[G_1]} ,\mathbf{0})$, and $(\dot{\mathbf{Z}}_i^{(t)})_{i\in[G_1]}$ are defined as follows
\begin{align}
    \dot{\mathbf{Z}}_i^{(t)}=&\begin{bmatrix}
        \ddot{\mathbf{Z}}^{(t)}_{i}\\
        \mathbf{H}^{(t)}_i\\
        \mathbf{0}_{(\Omega-X^{(t)}-|\mathcal{D}^{(t)}|)\times\gamma_{i}}
    \end{bmatrix},~\forall i\in[G_1^{(t)}],\label{eq:1-dzt1}\\
    \dot{\mathbf{Z}}^{(t)}_i=&\mathbf{0}_{\Omega\times\gamma_i},~\forall i\in[G_1^{(t)}+1:G_1],
\end{align}
where $(\ddot{\mathbf{Z}}^{(t)}_i)_{i\in[G_1^{(t)}]}$ are uniformly i.i.d. and independent of the increment $\boldsymbol{\Delta}^{(t)}$. In addition, since $\Omega=R_r-S-1$, and $|\mathcal{D}^{(t)}|\le R_r-K_c-X^{(t)}$, we have $\Omega-X^{(t)}-|\mathcal{D}^{(t)}|\geq K_c-S-1\geq 0$. Besides, $\mathbf{H}^{(t)}_{1}, \mathbf{H}^{(t)}_{2}, \cdots, \mathbf{H}^{(t)}_{G_1^{(t)}}$ are to be constructed such that $\mathbf{C}(n,:)\dot{\mathbf{M}}_1^{(t)}=\mathbf{0}$ for all $n\in\mathcal{D}^{(t)}$, whose existence is established as follows.

First, let us construct $\mathbf{H}^{(t)}_{1}$ as follows
\begin{align}
    &\mathbf{H}^{(t)}_{1}=-\mathbf{C}(\mathcal{D}^{(t)},[\alpha_1+X^{(t)}+1:\alpha_1+X^{(t)}+|\mathcal{D}^{(t)}|])^{-1}\times\notag\\
    &(\mathbf{C}(\mathcal{D}^{(t)},[\alpha_1])\dot{\mathbf{M}}^{(t)}_{1,1}([\alpha_1],:)
    +\notag\mathbf{C}(\mathcal{D}^{(t)},[\alpha_1+1:\alpha_1+X^{(t)}])\ddot{\mathbf{Z}}^{(t)}_{1}),
\end{align}
where according to Algorithm \ref{alg:scgen}, $\dot{\mathbf{M}}^{(t)}_{1,1}([\alpha_1],:)$ indeed represents the reshaped version of the increment vector $\boldsymbol{\Delta}^{(t)}$. It is evident to see that $\left(\mathbf{C}\dot{\mathbf{M}}_{1,1}^{(t)}\right)(\mathcal{D}^{(t)},:)=\mathbf{0}$ by our construction of $\mathbf{H}^{(t)}_{1}$. Now if $\mathbf{H}^{(t)}_{1},\mathbf{H}^{(t)}_{2},\cdots,\mathbf{H}^{(t)}_{i-1}$ are constructed such that for all $i'=1,2,\cdots,i-1$,
\begin{align}
    \left(\mathbf{C}\dot{\mathbf{M}}_{1,i'}^{(t)}\right)(\mathcal{D}^{(t)},:)=\mathbf{0}.
\end{align}
Then, let us construct $\mathbf{H}^{(t)}_{i}$ as follows
\begin{align}
    &\mathbf{H}^{(t)}_{i}=-\mathbf{C}(\mathcal{D}^{(t)},[\alpha_i+X^{(t)}+1:\alpha_i+X^{(t)}+|\mathcal{D}^{(t)}|])^{-1}\notag\\
    &\times(\mathbf{C}(\mathcal{D}^{(t)},[\alpha_i])\dot{\mathbf{M}}^{(t)}_{1,i}([\alpha_i],:)
    +\mathbf{C}(\mathcal{D}^{(t)},[\alpha_i+1:\alpha_i+X^{(t)}])\ddot{\mathbf{Z}}^{(t)}_{i}),
\end{align}
where similarly, according to Algorithm \ref{alg:scgen}, $\dot{\mathbf{M}}^{(t)}_{1,i}([\alpha_i],:)$ represents the reshaped version of the vector $\left[\dot{\mathbf{M}}^{(t)}_{1,1}(R_r+i-1,:),\dot{\mathbf{M}}^{(t)}_{1,2}(R_r+i-2,:),\cdots,\dot{\mathbf{M}}^{(t)}_{1,i-1}(R_r+1,:)\right]$, which is given as $\dot{\mathbf{M}}_{1,1}^{(t)},\dot{\mathbf{M}}_{1,2}^{(t)},\cdots,\dot{\mathbf{M}}_{1,i-1}^{(t)}$ are now fixed.
According to our construction, we can see that
\begin{align}
    &\left(\mathbf{C}\dot{\mathbf{M}}_{1,i}^{(t)}\right)(\mathcal{D}^{(t)},:)\notag\\
    =&\mathbf{C}(\mathcal{D}^{(t)},[\beta_i])\dot{\mathbf{M}}_{1,i}^{(t)}([\beta_i],:)\\
    =&\mathbf{C}(\mathcal{D}^{(t)},[\alpha_i])\dot{\mathbf{M}}^{(t)}_{1,i}([\alpha_i],:)
    +\mathbf{C}(\mathcal{D}^{(t)},[\alpha_i+1:\alpha_i+X^{(t)}])\ddot{\mathbf{Z}}^{(t)}_{i}\notag\\
    &+\mathbf{C}(\mathcal{D}^{(t)},[\alpha_i+X^{(t)}+1:\alpha_i+X^{(t)}+|\mathcal{D}^{(t)}|])\mathbf{H}^{(t)}_{i}\\
    =&\mathbf{0}.
\end{align}
In other words, the constructed $\mathbf{H}^{(t)}_{1},\mathbf{H}^{(t)}_{2},\cdots,\mathbf{H}^{(t)}_{i}$ now guarantee that for all $i'=1,2,\cdots, i$, $\left(\mathbf{C}\dot{\mathbf{M}}_{1,i'}^{(t)}\right)(\mathcal{D}^{(t)},:)=\mathbf{0}$. The existence of $\mathbf{H}^{(t)}_{1}, \mathbf{H}^{(t)}_{2}, \cdots, \mathbf{H}^{(t)}_{G_1^{(t)}}$ is thus concluded by induction on $i$.

Then, let us construct the coded increment $\mathbf{Q}^{(t)}_n, n\in[N]\setminus\mathcal{D}^{(t)}$ as follows
\begin{align}
    \mathbf{Q}^{(t)}_{n}=\mathbf{Q}^{(t)}_{1,n} &= \mathbf{C}(n,:)\dot{\mathbf{M}}_1^{(t)}.
\end{align}
Upon receiving the coded increment, Server $n, n\in[N]\setminus\mathcal{D}^{(t)}$ updates its storage according to $\mathbf{S}_n^{(t+1)}=\mathbf{S}^{(t)}_n+\mathbf{Q}^{(t)}_n$. The scheme is correct since according to the construction of $\mathbf{Q}^{(t)}_n, n\in[N]\setminus\mathcal{D}^{(t)}$, and the fact that $\mathbf{C}(n,:)\dot{\mathbf{M}}_1^{(t)}=\mathbf{0}$ for all $n\in\mathcal{D}^{(t)}$, upon the additive storage update, the storage structure is preserved. The scheme is $X^{(t)}$-secure since the increment symbols are protected by the MDS($N,X^{(t)}$)-coded noise symbols. For this setting, we have $\lambda_{G_1^{(t)}}=\frac{L}{\alpha_{G_1^{(t)}}}=\frac{L}{\alpha_1-G_1^{(t)}+1}=\frac{L}{R_r-|\mathcal{D}^{(t)}|-X^{(t)}}$, therefore, the normalized upload cost can be calculated as
\begin{align}
    C^{(t)}_u=\frac{(N-|\mathcal{D}^{(t)}|) \lambda_{G_1^{(t)}}}{L}=\frac{N-|\mathcal{D}^{(t)}|}{R_r-|\mathcal{D}^{(t)}|-X^{(t)}}.    
\end{align}
To show the optimality, consider the bound in \eqref{eq:defU-1}, averaging over all possible choices of $\mathcal{R}\subset[N]\setminus\mathcal{D}^{(t)},\mathcal{X}\subset\mathcal{R}$, such that $|\mathcal{X}|=X^{(t)},|\mathcal{R}|=R_r-|\mathcal{D}^{(t)}|$, we have 
\begin{align}
        \binom{N-|\mathcal{D}|-1}{R_r-|\mathcal{D}|-X^{(t)}-1}\sum_{n\in[N]\setminus\mathcal{D}}U_n&\geq\binom{N-|\mathcal{D}|}{R_r-|\mathcal{D}|-X^{(t)}}.
\end{align}
By expanding the binomial coefficients, we have
\begin{align}
    C_u^{(t)}&\ge\min_{(U_1, \cdots, U_N)\in\mathcal{U}}\sum_{n\in[N]\setminus\mathcal{D}^{(t)}}U_n \\
    &\geq \frac{N-|\mathcal{D}^{(t)}|}{R_r-|\mathcal{D}^{(t)}|-X^{(t)}},
\end{align}
which matches our achievability result.

\paragraph {When $|\mathcal{D}^{(t)}|+X^{(t)}>R_r-K_c$}
Define $P^{(t)}=|\mathcal{D}^{(t)}|+X^{(t)}+K_c-R_r$ and {$G_2^{(t)}=N-R_r-\Omega+1+|\mathcal{D}_1^{(t)}|+X^{(t)}$}. Let $\dot{\mathbf{M}}_1^{(t)},\dot{\mathbf{M}}_2^{(t)}=\FuncSty{PSCGen}(\mathbf{\Delta}^{(t)},(\dot{\mathbf{Z}}_i^{(t)})_{i\in[G_1]}, (\dot{\mathbf{Z}}_{i,j}^{(t)})_{i\in[P^{(t)}],j\in[G_2^{(t)}]})$. Besides, $(\dot{\mathbf{Z}}_i^{(t)})_{i\in[G_1]})$ are defined as follows
\begin{align}
    \dot{\mathbf{Z}}_i^{(t)}=\begin{bmatrix}
        \ddot{\mathbf{Z}}^{(t)}_{i}\\
        \mathbf{H}^{(t)}_i\\
        \mathbf{0}_{(\Omega-X^{(t)}-|\mathcal{D}^{(t)}|)\times\gamma_{i}}
    \end{bmatrix},&~\forall i\in[G_1],
\end{align}
where $(\ddot{\mathbf{Z}}^{(t)}_i)_{i\in[G_1]}$ are uniformly i.i.d. over $\mathbb{F}_q^{X^{(t)}\times(\gamma_{1}+\gamma_{2}+\cdots+\gamma_{G_1})}$ and independent of the increment $\boldsymbol{\Delta}^{(t)}$, and $\mathbf{H}^{(t)}_{1}, \mathbf{H}^{(t)}_{2}, \cdots, \mathbf{H}^{(t)}_{G_1}$ are to be constructed such that $\mathbf{C}(n,:)\dot{\mathbf{M}}^{(t)}=\mathbf{0}$ for all $n\in\mathcal{D}^{(t)}$ following a strategy similar to that described in the preceding subsection. Finally, $(\dot{\mathbf{Z}}_{i,j}^{(t)})_{i\in[P^{(t)}],j\in[G_2^{(t)}]}$ are defined as follows
\begin{align}
    \dot{\mathbf{Z}}_{i,j}^{(t)}=&\begin{bmatrix}
        \ddot{\mathbf{Z}}^{(t)}_{i,j}\\
        \mathbf{H}^{(t)}_{i,j}\\
        \mathbf{0}_{(\Omega-X^{(t)}-|\mathcal{D}_1^{(t)}|)\times\gamma'_{j}}
    \end{bmatrix},~\forall {i\in[P^{(t)}],j\in[G_2^{(t)}]},\\
    \dot{\mathbf{Z}}^{(t)}_{i,j}=&\mathbf{0}_{\Omega\times\gamma_j},~\forall i\in[P^{(t)}+1:P],j\in[G_2^{(t)}+1:G_2],
\end{align}
where $(\ddot{\mathbf{Z}}^{(t)}_{i,j})_{i\in[P^{(t)}],j\in[G_2^{(t)}]}$ are uniformly i.i.d. and independent of the increment $\boldsymbol{\Delta}^{(t)}$, and $\mathbf{H}^{(t)}_{1,1}, \cdots, \mathbf{H}^{(t)}_{1,G_2^{(t)}},\allowbreak\cdots,\mathbf{H}^{(t)}_{P^{(t)},G_2^{(t)}}$ are to be constructed such that $\mathbf{C}(n,:)\dot{\mathbf{M}}_2^{(t)}=\mathbf{0}$ for all $n\in\mathcal{D}_1^{(t)}$. Now we are ready to construct the coded increment $\mathbf{Q}^{(t)}_n, n\in[N]\setminus\mathcal{D}^{(t)}$ as follows
\begin{align}
    &\mathbf{Q}^{(t)}_{1,n} = \mathbf{C}(n,:)\dot{\mathbf{M}}_1^{(t)},&&\forall n \in [N]\setminus\mathcal{D}^{(t)},\\
    &\mathbf{Q}^{(t)}_{2,n} = \mathbf{C}(n,N-S)\dot{\mathbf{M}}_2^{(t)},&&\forall n \in [S+1:N]\setminus\mathcal{D}^{(t)},\\
    &\mathbf{Q}^{(t)}_{n}=\mathbf{Q}^{(t)}_{1,n},&&\forall n\in [S]\setminus\mathcal{D}^{(t)},\\
        &\mathbf{Q}^{(t)}_{n}=(\mathbf{Q}^{(t)}_{1,n},\mathbf{Q}^{(t)}_{2,n}),&&\forall n\in [S+1:N]\setminus\mathcal{D}^{(t)}.
\end{align}
Upon receiving the coded increment, server $n, n\in[N]\setminus\mathcal{D}^{(t)}$ updates its storage according to $\mathbf{S}_n^{(t+1)}=\mathbf{S}^{(t)}_n+\mathbf{Q}^{(t)}_n$. Correctness and $X^{(t)}$-security of the scheme can be similarly verified. For this setting, $\lambda'_{G_2^{(t)}}=\frac{L'}{\alpha'_{G_2^{(t)}}}=\frac{L}{K_c(N-S-\Omega-G_2^{(t)}+1)}=\frac{L}{K_c(R_r-S-|\mathcal{D}_1^{(t)}|-X^{(t)})}$, hence the normalized upload cost is
\begin{align}
    &C_u^{(t)}=\frac{S-|\mathcal{D}_2^{(t)}|}{K_c}+(N-S-|\mathcal{D}_1^{(t)}|)(\frac{1}{K_c}+\frac{P^{(t)}\lambda'_{G_2^{(t)}}}{L}) \\
    &=\frac{S-|\mathcal{D}_2^{(t)}|}{K_c}+\frac{(N-S-|\mathcal{D}_1^{(t)}|)(K_c-S+|\mathcal{D}_2^{(t)}|)}{K_c(R_r-S-|\mathcal{D}_1^{(t)}|-X^{(t)})} \\
    &=\frac{N-S-|D_1^{(t)}|}{R_r-S-|D_1^{(t)}|-X^{(t)}}-\frac{(S-|D_2^{(t)}|)(N-R_r+X^{(t)})}{K_c(R_r-S-|D_1^{(t)}|-X^{(t)})}.
\end{align}
To show the optimality, consider the bound \eqref{eq:defU-1}, averaging over all possible choices of $\mathcal{R}\subset[N]\setminus\mathcal{D}^{(t)},\mathcal{X}\subset\mathcal{R}$, such that $|\mathcal{X}|=X^{(t)},|\mathcal{R}|=R_r-|\mathcal{D}^{(t)}|,\mathcal{R}\setminus\mathcal{X}\supset[S]\setminus\mathcal{D}^{(t)}$, we have
\begin{align}
    \binom{N-S-|\mathcal{D}^{(t)}_1|-1}{R_r-S-|\mathcal{D}^{(t)}_1|-X^{(t)}-1}\sum_{n\in[S+1:N]\setminus\mathcal{D}^{(t)}}U_n+&
    \binom{N-S-|\mathcal{D}^{(t)}_1|}{R_r-S-|\mathcal{D}^{(t)}_1|-X^{(t)}}\sum_{n\in[S]\setminus\mathcal{D}^{(t)}}U_n \nonumber \\
    &\geq\binom{N-S-|\mathcal{D}^{(t)}_1|}{R_r-S-|\mathcal{D}^{(t)}_1|-X^{(t)}},\\
    \Rightarrow\sum_{n\in[S+1:N]\setminus\mathcal{D}^{(t)}}U_n+\frac{N-S-|\mathcal{D}^{(t)}_1|}{R_r-S-|\mathcal{D}^{(t)}_1|-X^{(t)}}&\sum_{n\in[S]\setminus\mathcal{D}^{(t)}}U_n
    \geq\frac{N-S-|\mathcal{D}^{(t)}_1|}{R_r-S-|\mathcal{D}^{(t)}_1|-X^{(t)}}.\label{lp-u}
\end{align}
In addition, according to the bound \eqref{eq:defU}, we have
\begin{align}
    -\frac{X^{(t)}}{R_r-S-|\mathcal{D}^{(t)}_1|-X^{(t)}}\sum_{n\in[S]\setminus\mathcal{D}^{(t)}}U_n\geq-\frac{(S-|\mathcal{D}^{(t)}_2|)X^{(t)}}{K_c(R_r-S-|\mathcal{D}^{(t)}_1|-X^{(t)})}\label{lp-d2}.
\end{align}
Combining the above bounds, we have
\begin{align}
     &C_u^{(t)}\ge \min_{(U_1, \cdots, U_N)\in\mathcal{U}}\sum_{n\in[N]\setminus\mathcal{D}^{(t)}}U_n\\
    \geq&\frac{N-S-|\mathcal{D}^{(t)}_1|}{R_r-S-|\mathcal{D}^{(t)}_1|-X^{(t)}}-\frac{(S-|\mathcal{D}^{(t)}_2|)X^{(t)}}{K_c(R_r-S-|\mathcal{D}^{(t)}_1|-X^{(t)})}. \label{lp-u-bound2}
\end{align}
Therefore the normalized upload cost of our scheme achieves the lower bound.

\section{Conclusion}
In this paper, we explored the problem of RDCDS with partially storage constrained servers and completely characterized its fundamental limits, i.e., the minimum number of available servers for feasible update operation and the minimum communication costs for read and update operations. These results can be viewed as our first attempt towards the settlement of RDCDS with heterogeneous storage where the capacity of each server's storage is arbitrarily constrained. Also, our results are of interest to the problem of private read and write \cite{jia2024read} with storage constraints, where the read and update operations are required to be private, i.e., disclose no information on the index of the message being processed.

\bibliography{ref}
\bibliographystyle{IEEEtran}

\end{document}